\documentclass[11pt]{article}

\usepackage{amssymb,amsmath,amsthm} 
\usepackage[margin=1in]{geometry}
\usepackage{graphicx,ctable,booktabs}
\usepackage{verbatim}
\usepackage{enumerate}
\usepackage{float}
\usepackage{epstopdf}
\usepackage{multirow}
\usepackage{color}

\usepackage{natbib}
\setcitestyle{authoryear}

\usepackage{lineno}

\begin{document}

\title{A Fluid-Structure Interaction Model of the \\ Zebrafish Aortic Valve}
\author{ 
Alexander D. Kaiser, PhD $^{1,2}$ 
Jing Wang, MS$^{3}$  \\
Aaron L. Brown, MS$^{2,4}$ 
Enbo Zhu, PhD$^{5}$  \\
Tzung Hsiai, MD PhD$^{3,5}$   
\&
Alison L. Marsden, PhD$^{2,6,7,8}$
}

\date{\today \\
\vspace{0.5cm}
    {\small
    $^{1}$Department of Cardiothoracic Surgery, Stanford University, Stanford CA;  \\ 
    $^{2}$Stanford Cardiovascular Institute, Stanford CA; \\
    $^{3}$Department of Bioengineering, University of California, Los Angeles, Los Angeles, CA; \\
    $^{4}$Department of Mechanical Engineering, Stanford University, Stanford CA;  \\
    $^{5}$Division of Cardiology, Department of Medicine, School of Medicine, University of California, Los Angeles, Los Angeles, CA; \\ 
    $^{6}$Department of Pediatrics (Cardiology), Stanford University, Stanford CA; \\
    $^{7}$Department of Bioengineering, Stanford University, Stanford CA; \\
    $^{8}$Institute for Computational and Mathematical Engineering, Stanford University, Stanford CA \\
    }
}

\maketitle

\thispagestyle{empty}

\begin{abstract}

The zebrafish is a valuable model organism for studying cardiac development and diseases due to its many shared aspects of genetics and anatomy with humans and ease of experimental manipulations.
Computational fluid-structure interaction (FSI) simulations are an efficient and highly controllable means to study the function of cardiac valves in development and diseases. 
Due to their small scales, little is known about the mechanical properties of zebrafish cardiac valves, limiting existing computational studies of zebrafish aortic valves and their interaction with blood. 
To circumvent these limitations, we took a largely first-principles approach called \emph{design-based elasticity} that allows us to derive valve geometry, fiber orientation and material properties. 
In FSI simulations of an adult zebrafish aortic valve, these models produce realistic flow rates when driven by physiological pressures and demonstrate the spatiotemporal dynamics of valvular mechanical properties. 
These models can be used for future studies of zebrafish cardiac hemodynamics, development, and disease.

\end{abstract}

\section{Introduction}

The zebrafish is a widely used model organism to study cardiac development and diseases, sharing 82\% of disease-associated genes with humans, including 96\% of dilated cardiomyopathy (DCM) genes \citep{howe2013zebrafish,shih2015cardiac}. 
The zebrafish provides unique advantages in high-throughput experiments via its translucent body during the embryonic stage that permits \emph{in vivo} imaging, short gestational period and time to maturity. 
The zebrafish has a two-chambered heart with two valves, in contrast to four chambers with four valves in humans. 
The heart possesses one atrium, an atrioventricular (AV) valve, one ventricle, an outflow structure called the bulbus arteriosus, and a semilunar aortic (outflow tract or OFT) valve in the outflow tract.
The aortic valve is bicuspid, with two leaflets of approximately equal size and shape. 
This valve resembles a rare congenitally diseased phenotype that occurs in humans, a Sievers Type 0 bicuspid valve that has no raphe \citep{sievers2007classification}. 
Computational fluid dynamics (CFD), including fluid-structure interaction (FSI) simulations, has facilitated uncovering novel mechanisms underlying the zebrafish cardiac morphogenesis \citep{salman2020advanced}. 
While the FSI simulations of endocardial wall elucidated shear-stress mediated valvular development, successful attempts to simulate valve-fluid interaction in zebrafish have not been documented \citep{hsu2019contractile}.
Moreover, the mechanical properties of the zebrafish aortic valve are poorly understood and difficult to directly measure because of its small size.  

In this work, we constructed a computational model of an adult zebrafish aortic valve and conducted FSI simulations to model its interaction with blood. 
Due to the lack of information on the gross morphology of the valve and its material properties, as well as our previous successes with the method, we applied a nearly first-principles approach that we refer to as \emph{design-based elasticity}. 
In this approach, we derived a system of partial differential equations to represent the closed valve as it supports a pressure load.  
The solution to these equations was then used to derive the material properties, fiber orientation, and reference configuration of the model valve.  
By tuning parameters in these equations, we designed the valve to achieve realistic flow rates and closure when driven by physiological pressures. 

Initially developed to model the human mitral valve \citep{kaiser2019modeling}, we have previously applied design-based elasticity to the human aortic valve \citep{kaiser2020designbased}. We studied the effect of phenotype on hemodynamics in Sievers Type 1 bicuspid valves \citep{kaiser2022controlled} and showed that model flows compared well to an \emph{in vitro} experiment measured with flow with 4D flow Magnetic Resonance Imaging (4DMRI) \citep{kaiser2023comparison}. 
We conducted initial simulation-guided design studies of bicuspidization repair of the aortic valve \citep{kaiser2023simulation} and validated predicted trends in pressure gradient in vitro \citep{choi2024combined,choi2024effect}.
Our methods have thus proven effective for modeling a variety of flows involving heart valves and compare well to experimental data, and thus are appropriate for modeling the zebrafish aortic valve.

The zebrafish aortic valve has Reynolds number order 1 (Sec. \ref{results}), indicating an intermediate regime, neither stokes flow nor fully inertial. 
In contrast to the zebrafish, the human aortic valve operates in an inertial regime with Reynolds number $>10^3$ \citep{kaiser2022controlled}. 
In this different flow environment, the human aortic valve has been extensively studied with FSI simulations. 
Many studies use the immersed boundary method \citep{ib_acta_numerica}, as in this study, and its extensions \citep{kamensky2015immersogeometric}. 
Studies have examined 
jet analysis with Lagrangian coherent structures \citep{shadden2010computational}, 
hemodynamics associated with bicuspid aortic valve, \citep{marom2013fully,emendi2021patient,kaiser2022controlled}
and 
the interaction of aortic root and leaflet geometry in valve repair \citep{marom2013aortic}. 
Studies on prosthetic medical devices have 
modeled dynamics of valves in an \emph{in vitro} pulse duplicator \citep{lee2020fluid}, 
examined leaflet flutter \citep{lee2021bioprosthetic}, 
revealed mechanisms of flow instabilities leading to turbulence \citep{bornemann2024instability} and assessed hemodynamics for thrombosis potential \citep{bornemann2024relation}.

% add some more about overall zebrafish cv physiology 
% probably should add some zebrafish flow studies here 
In the zebrafish, other FSI studies have examined flow in the developing heart \citep{battista2018fluid,hsu2019contractile,lee2013moving,lee2018spatial,vedula2017method}, including one study with simplified two-dimensional endocardial cushions that develop into valves \citep{miller2011fluid}, and adult vasculature \citep{van2023fluid}.

To our knowledge this is the first study to perform three-dimensional FSI on the zebrafish aortic valve. 
Our model produced realistic flow rates and robust closure in diastole over two cardiac cycles. 
This study serves as a proof of concept towards future studies of the mechanisms of zebrafish aortic valve function, disease and development. 
Code for this project is open source and freely available (github.com/alexkaiser/heart\_valves).

\section{Methods}

\subsection{Imaging and Anatomy}

\begin{figure}[t]
\includegraphics[width=\textwidth]{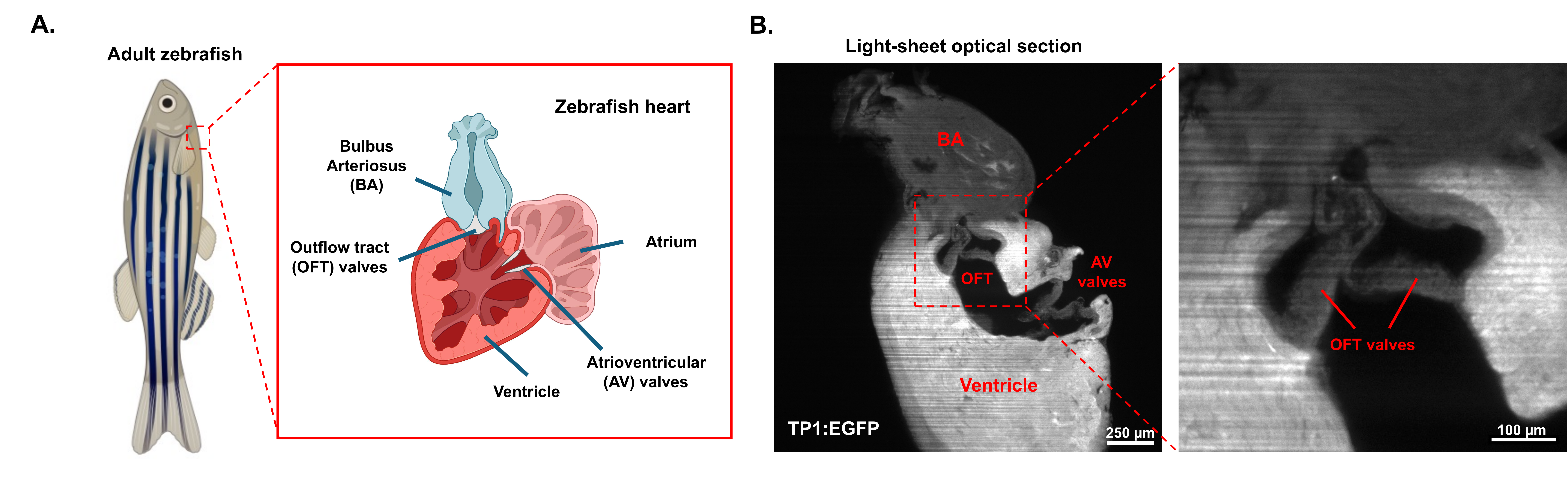}
\caption{ Anatomy of zebrafish heart and light-sheet optical section of OFT (aortic) valves. 
A. Illustration of the adult zebrafish and its cardiac structure.
B. Light-sheet imaging of a cleared 1 year of age zebrafish heart showing the AV and OFT leaflets.}
\label{fish_lightsheet}
\end{figure}

A schematic together with light-sheet images of the zebrafish heart is shown in Figure \ref{fish_lightsheet}. 
The clearing of the 1 year-old Tg(TP1:EGFP) \citep{parsons2009notch} zebrafish heart was carried out following a CUBIC-based protocol from Tokyo Chemical Industry (TCI) \citep{tainaka2018chemical}. 
Briefly, the heart was washed in PBS before undergoing serial incubations in CUBIC-L (TCI, T3740) and CUBIC-R+(M) (TCI, T3741) solutions at room temperature. 
Finally, the cleared heart was immersed in the mounting solution (TCI, M3294, RI = 1.520) and imaged using a customized light-sheet system at 6.3X magnification. Adult zebrafish were raised and bred in the UCLA Zebrafish Core Facility according to standard protocols \citep{westerfield2007zebrafish}. All experiments with zebrafish were performed in compliance with and with the approval of a UCLA Institutional Animal Care and Use Committee protocol (ID: ARC-2015-055).

\subsection{Model construction}

Models were constructed with a nearly first-principles method called design-based elasticity that we previously applied to model the human aortic valve \citep{kaiser2020designbased}. 
In overview, to construct the model, we specified that tension in the leaflets must support a pressure and derived a resulting system of partial differential equations. 
Solving these equations yielded the predicted loaded configuration of the valve, with parameters and boundary conditions tuned to design the desired leaflet geometry. 
By temporarily prescribing a uniform pressure load in the model construction process, the necessary material properties to close under dynamic pressure loading in FSI simulations can be \emph{derived}. 
Further, this method produces heterogeneous and anisotropic material properties and fiber orientations. 
This contrasts with standard techniques, in which material properties would be measured and prescribed from experimental data, which are highly challenging to obtain for a zebrafish. 

The steps of this method are summarized as follows with changes for the zebrafish highlighted. 
First, we temporarily assumed that tension in the leaflets supported a static, uniform pressure load $p$.
The leaflet was represented as an unknown parametric surface $ \mathbf X(u,v) : \Omega \subset \mathbb R^{2} \to \mathbb R^{3}$. Curves $\mathbf X(u,v = v_{0})$ on which $u$ varies for any given, fixed $v_{0}$ run circumferentially, corresponding to the anatomical fiber direction, and exerted tension $S$. 
Similarly, curves $\mathbf X(u = u_{0},v)$ on which $v$ varies for any fixed $u_{0}$ run radially, corresponding to the cross-fiber direction, and exerted tension $T$.
We specified that tension on the boundary of an arbitrary patch of leaflet balanced pressure, applied the fundamental theorem of calculus to convert boundary terms to surface terms, then finally dropped the integrals as the region of interest was arbitrary. 
Let single bars, $| \cdot |$, denote the Euclidean norm and subscripts denote partial derivatives.
We thus obtained the following system of partial differential equations that represent the predicted, loaded configuration of the leaflets,
\begin{align} 
0 = p  (  \mathbf X_{u} \times \mathbf X_{v} )  +   \frac{\partial}{\partial u}  \left( S \frac{ \mathbf X_{u} }{ |\mathbf X_{u}| } \right)  +  \frac{\partial}{\partial v}  \left( T \frac{ \mathbf X_{v} }{|\mathbf X_{v}|} \right).    
\label{eq_eqns}
\end{align}
We then assigned the following tension law for $S$ and $T$, 
\begin{align}
S(u,v) &= \alpha \left( 1 - \frac{1}{1 + |\mathbf X_{u}|^{2} / a^{2} } \right), \quad
T(u,v) = \beta \left( 1 - \frac{1}{1 + |\mathbf X_{v}|^{2} / b^{2} } \right), \label{dec_tension}  
\end{align}
where $\alpha$ and $\beta$ control the maximum tension in each material direction, and $a$ and $b$ control spacing of the mathematical fibers in the model. 
Expression \eqref{dec_tension} acts, colloquially, as a ``mathematical search function,'' that allowed the solution to find a heterogeneous field of tensions that support the specified load. 
Non-uniform pressure or shear forces could be applied to the leaflets in this model construction step were their precise values available, but we expect the pressure to be close to uniform in FSI simulations and shear forces to be much smaller than pressure forces when the valve is closed.  
Note that in FSI simulations, the loading on the valve will include pressure, which is expected to be approximately uniform with minor spatial variation, and other fluid forces determined by their coupled interaction.

To design a realistic model for a zebrafish, the pressure $p$ was set to 0.92 mmHg, the mean pressure difference across the valve when the aortic pressure was greater than the ventricular pressure \citep{hu2001cardiac}. 
The model was set to have two symmetrical leaflets, each subtending half the annular circumference.
The valve diameter was set to 363 microns and commissure height to 285 microns, based on measurements from a light-sheet image of the valve (Fig. \ref{fish_lightsheet}B). 
This imaging modality allowed viewing of the annulus, but was highly collapsed and compressed since the fish was deceased at the time of imaging. 
Thus, these are low estimates on the valve radius and height, but were used due to lack of comparable \emph{in vivo} data.  
The parameters that control the maximum tension were set to $\alpha = 4.1 \cdot 10^{5}$ and  
$\beta = 1.4 \cdot 10^{4}$ dynes. 
The spacing parameters were taken to be dimensionless and $a$ varied linearly between $a = 26.2-100$ from the annulus to the free edge, while $b = 40$ throughout the leaflet.
The leaflet mesh was constructed to have a mesh spacing of approximately 2.5 microns, half that of the fluid mesh. To achieve this, 384 points were placed along the annulus and 65 points in the radial direction.  
The resulting mesh had mean spacing 3.0 microns circumferentially and 1.6 microns radially at rest. 
The mesh spacing changed dynamically in the FSI simulations and was typically larger radially due to radial stretch through the cardiac cycle.
These parameters were designed by trial and error to produce a model free edge length (leaflet width) of 571 microns in the reference configuration, or approximately 1.57 times the annular diameter, as in previous work on bicuspidization repair of the aortic valve \citep{kaiser2023simulation}. 
The resultant leaflet height (or geometric height) was 169 microns in the reference configuration.

To derive a constitutive law from the predicted, loaded configuration, we prescribed uniform stretch of $\lambda_{c} = 1.15$ and  $\lambda_{r} = 1.54$ circumferentially and radially \citep{yap2009dynamic}. 
We derived the reference length $R$ for all links in the discretized model via solving the equation $\lambda = L/R$, where $\lambda$ is the relevant stretch and $L$ is the current length of a given edge or ``link'' in the discretized model. 
The stretch-tension response of each link was modeled with an exponential function
\begin{align}
\tau = \kappa (e^{\eta (\lambda - 1)} - 1)
\end{align}
with exponential rates $\eta$ = 57.46 circumferentially and $\eta$ = 22.40 radially based on experimental data \citep{may2009hyperelastic}. 
The stiffness of each link $\kappa$ was then scaled to take the tension required in the solution of the equilibrium equations \eqref{eq_eqns} at the prescribed stretches. 
This process resulted in anisotropic, heterogeneous material properties. 
The resultant mean tangent moduli circumferentially and radially were $2.83 \cdot 10^{6}$ and $1.16 \cdot 10^{5}$ dynes/cm$^{2}$ respectively at the predicted loaded stretch ratios. 
The emergent stretches and tangent moduli in each direction were comparable but slightly lower in the simulation (See Sec. \ref{results} and Fig. \ref{stretch_plots}).

To obtain an open initial configuration suitable for FSI simulations, an additional equilibrium equation was solved with $p = 0$ mmHg and the constitutive law just derived. 
Finally, two additional membranous layers were added to create a leaflet thickness of 6.3 microns, which was scaled proportionally to valve radius from our previous models \citep{kaiser2020designbased}. 
(Note that image resolution in the out of plane direction may cause the leaflets in Fig. \ref{fish_lightsheet}B to appear artificially thicker than they are, and thus thickness was not assessed from this image.)
This thickness was meant to model a nearly-zero-thickness, membranous structure.  
Previously, we found that a finite thickness of three layers spaced half the fluid mesh width apart, one fluid mesh width in total, mitigated the ``grid-aligned artifact'' that occurred when an immersed boundary supports a jump in pressure \citep{kaiser2019modeling}. 
The layers were attached to each other with stiff linear springs with spring constant 78.90 dynes/cm, which was determined to be the stiffest value that did not result in further time step restrictions.
The stiffnesses of each membrane layer were set to one third of the initial stiffnesses across the three layers.  
The model was fully fiber-based, in that all circumferential, radial and cross-layer forces are computed along fibers, curves that exert tension along their axis. 
Links that were longer than twice the fluid mesh width were split and assigned the same relative spring constant. 
Stretch, stress and tangent modulus from FSI simulations were computed from the original membranous layer on the aortic side of the leaflet.

\subsection{Fluid-Structure Interaction}

Fluid-structure interaction simulations were performed with the Immersed Boundary Method \citep{ib_acta_numerica} in the open-source solver IBAMR (Immersed Boundary Adaptive Mesh Refinement) \citep{griffith2010parallel}.
The fluid had a density of 1.0 g/ml and viscosity 0.04 poise. 
The spatial resolution was set to 5 microns and the time step was set to $5 \cdot 10^{-7}$ s. 
The valve was placed in a cylindrical test chamber. 
Two cardiac cycles were simulated and results from the second cycle were analyzed. 
At the inlet, a pressure waveform from experimental measurements with heart rate 185 beats per minute was prescribed \citep{hu2001cardiac}. 
At the outlet, pressure boundary conditions were determined by an RCR (resistor capacitor resistor) lumped parameter network tuned for minimum, maximum and mean aortic pressure of 0.89, 2.08 and 1.39 mmHg respectively \citep{hu2001cardiac} and stroke volume of 266 nl \citep{van2023fluid}. 
Additional details of the FSI setup and a mesh resolution study are shown in the Appendix.

\section{Results}
\label{results}

\begin{figure}[th]
\setlength{\tabcolsep}{0.0pt}
% \hfill \includegraphics[width=.08\textwidth]{colorbar_zebrafish.jpeg}

\vspace{5pt}
\begin{tabular}{ c | c | c | c | c | c | c | c | c | c  }
% \begin{tabular}{ c  c  c  c  c  c  c  c  c}
% \hspace{-40pt}
\includegraphics[width=.10\textwidth]{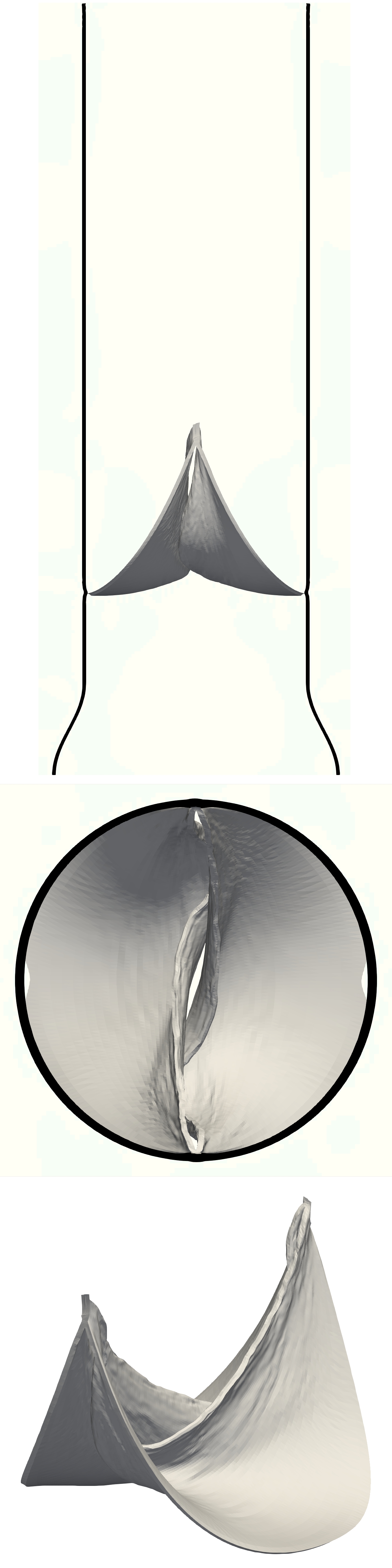} & 
\includegraphics[width=.10\textwidth]{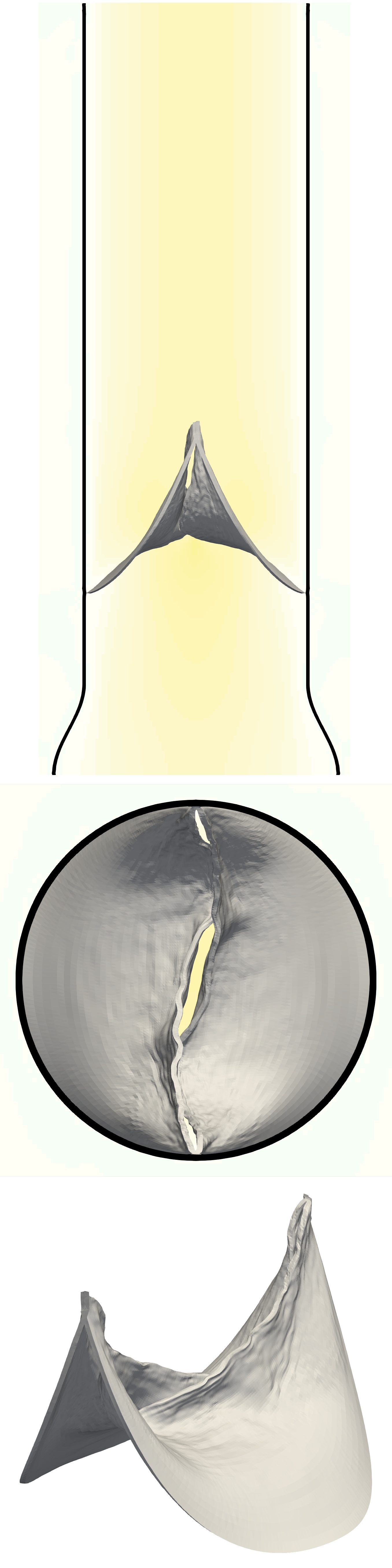} &
\includegraphics[width=.10\textwidth]{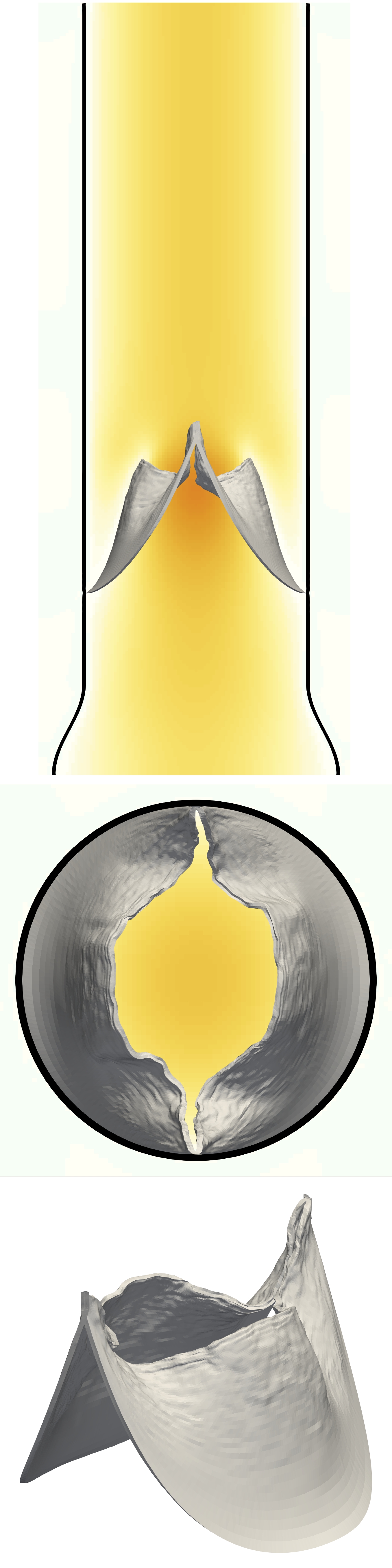} & 
\includegraphics[width=.10\textwidth]{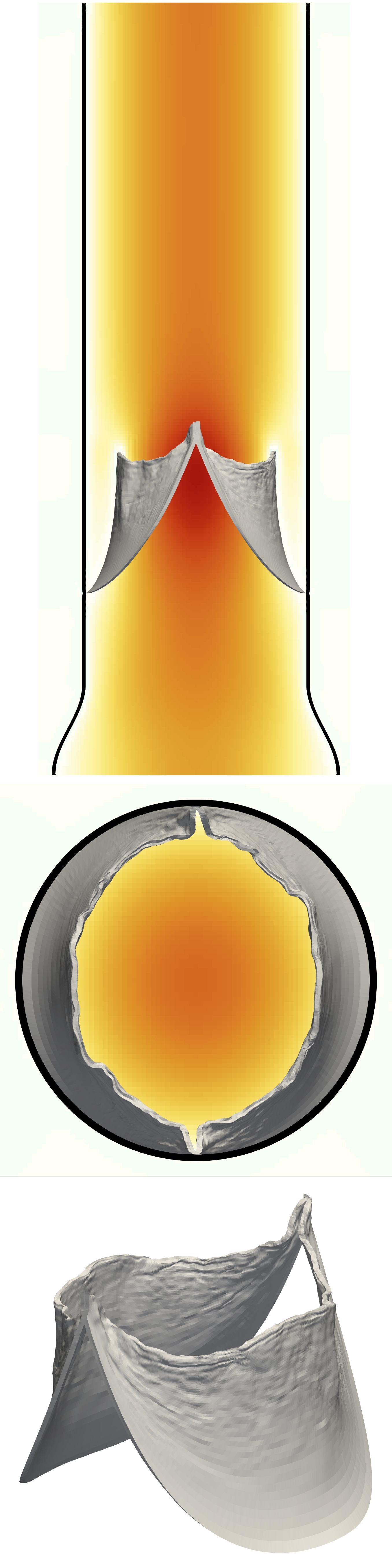} &
\includegraphics[width=.10\textwidth]{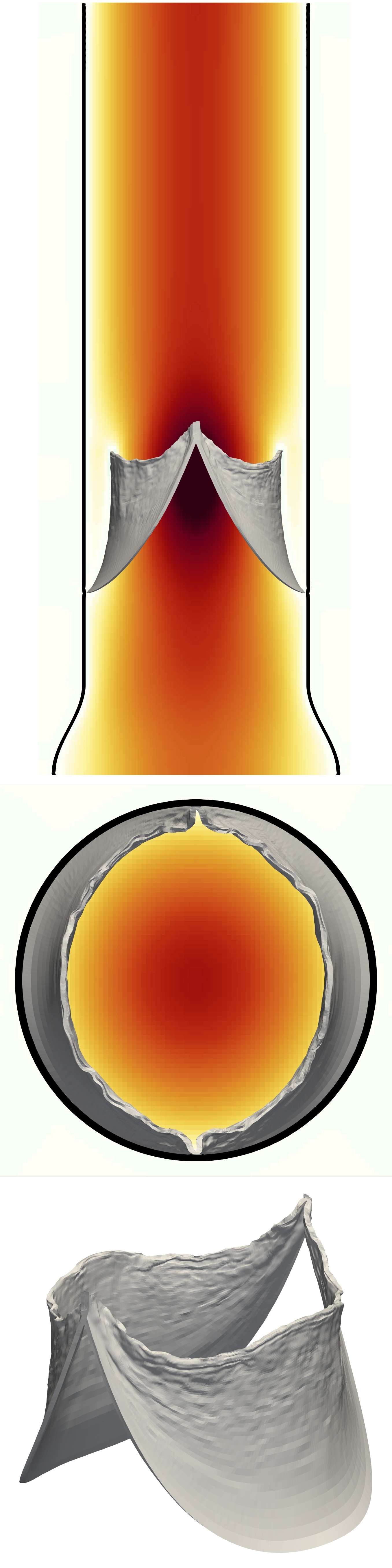} & 
\includegraphics[width=.10\textwidth]{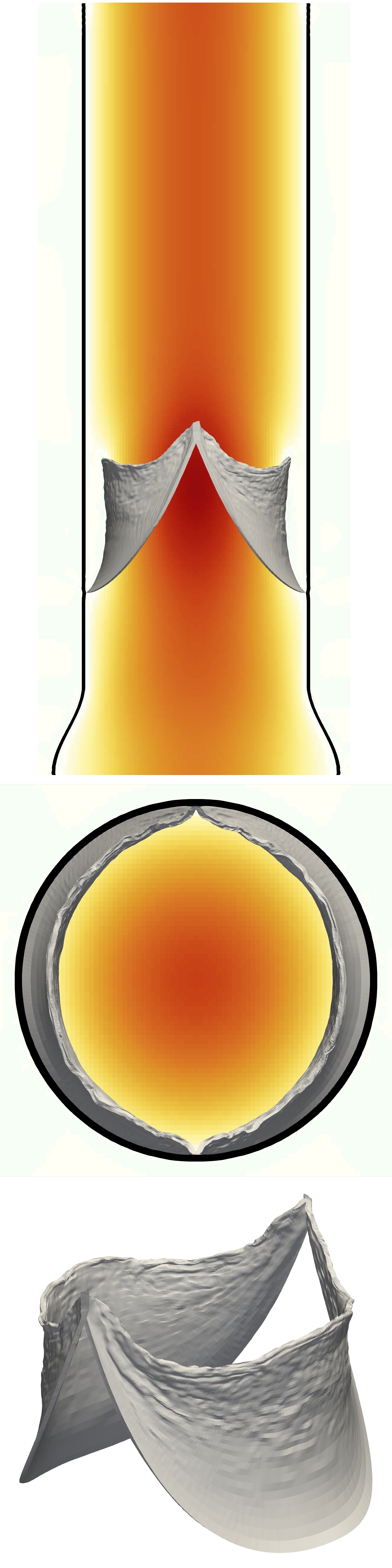} &
\includegraphics[width=.10\textwidth]{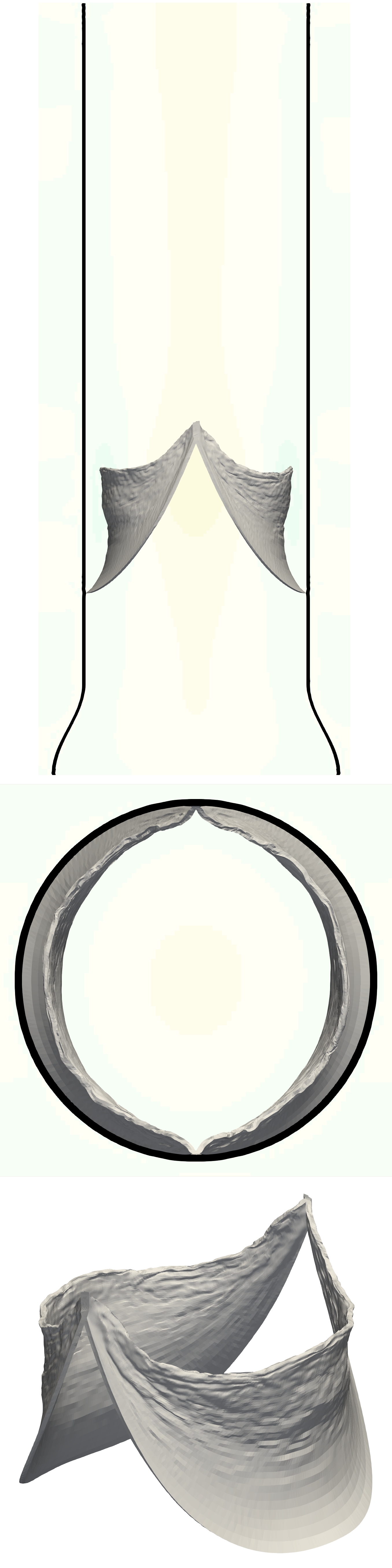} & 
\includegraphics[width=.10\textwidth]{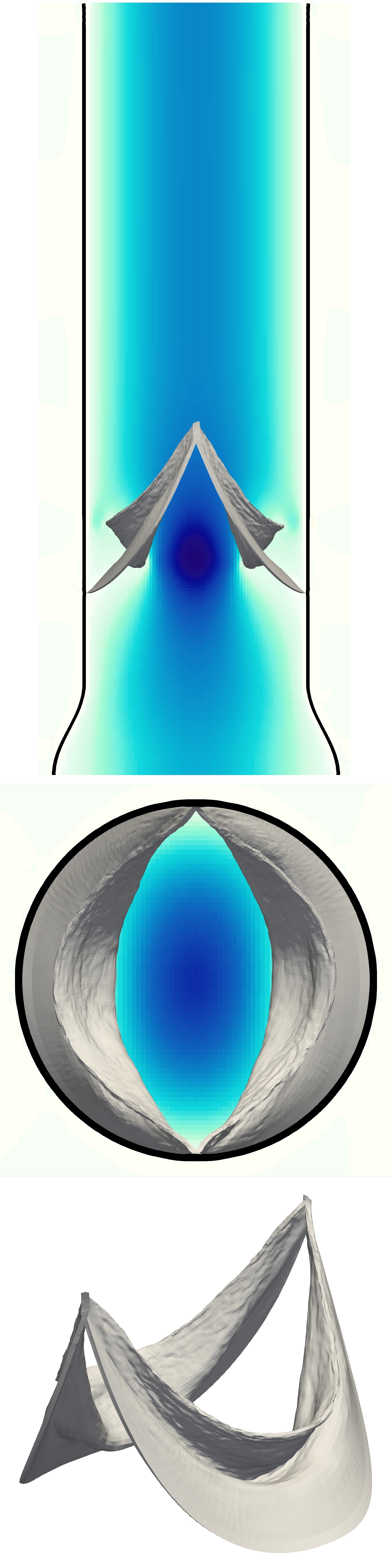} & 
\includegraphics[width=.10\textwidth]{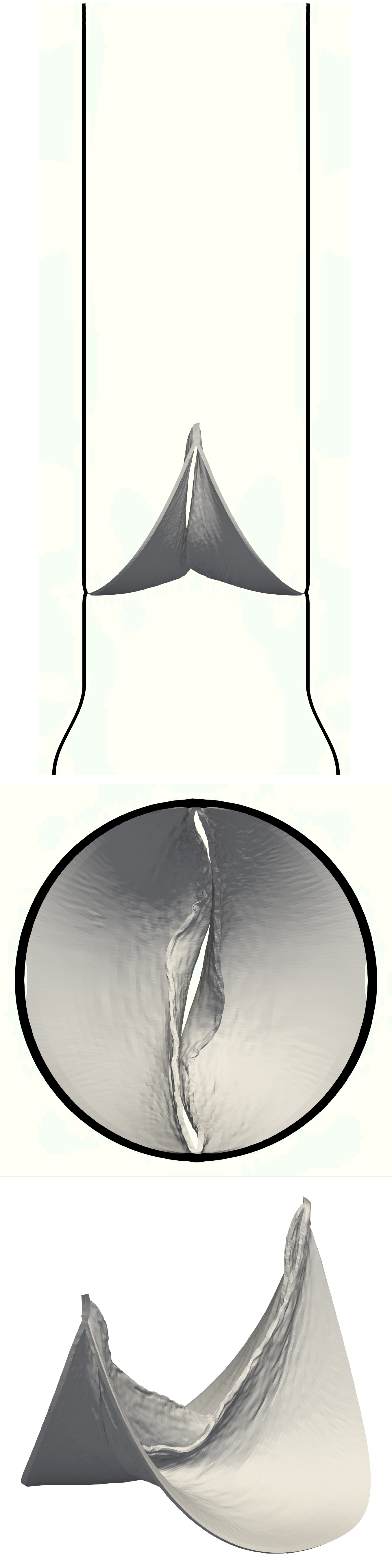} & 
\raisebox{35mm}{\includegraphics[width=.08\textwidth]{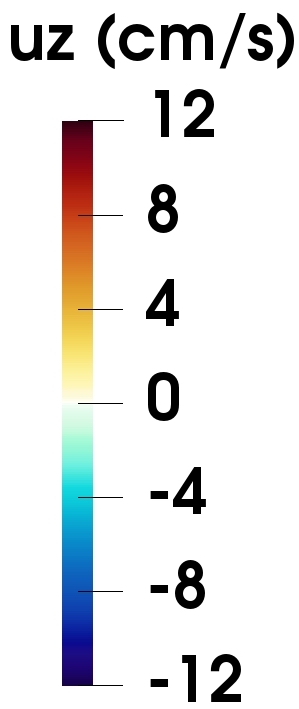}}
\\ 
\end{tabular}
\caption{Flows through the cardiac cycle, depicting a slice view of the vertical component of velocity $u_{z}$ with a side view of the valve (top row), the valve viewed \emph{en face} with a slice view of the vertical component of velocity at the aortic annulus (middle row) and the valve viewed at an angle (bottom row). 
The penultimate frame shows the closing transient, followed by closure without regurgitation in the final frame.
}
\label{cycle_flow}
\end{figure}

The valve shows unrestricted forward flow during systole and robust closure during diastole over two cardiac cycles (Fig. \ref{cycle_flow}). 
The velocity field showed no evidence of regurgitation and low magnitude flow during closure. 
Twist was visible in the closed configuration, indicating presence of excess free edge length during diastole. 
The leaflets showed no billow or prolapse. 
The valve opens to nearly the radius of the cylindrical test chamber, indicating sufficient free edge length for maximal orifice area during systole. 
The forward flow is highly laminar, without any appearance of vortices or instabilities in the flow. 
When the flow rate nears zero, the local velocity is also low-magnitude while the valve remains open. 
The closing transient appears roughly symmetric, with a transient crease in both leaflets. 
Finally, the valve again fully closes with slight twist and no appearance of regurgitation. 
The flow dynamics are further depicted in the movie included with the supplemental information. 

\begin{figure}[th]
\setlength{\tabcolsep}{0.0pt}
\vspace{5pt}
%\begin{tabular}{ c | c | c | c | c | c | c | c | c | c  }
\begin{tabular}{ c  c  c  c  c  c  c  c  c c  }
\includegraphics[width=.10\textwidth]{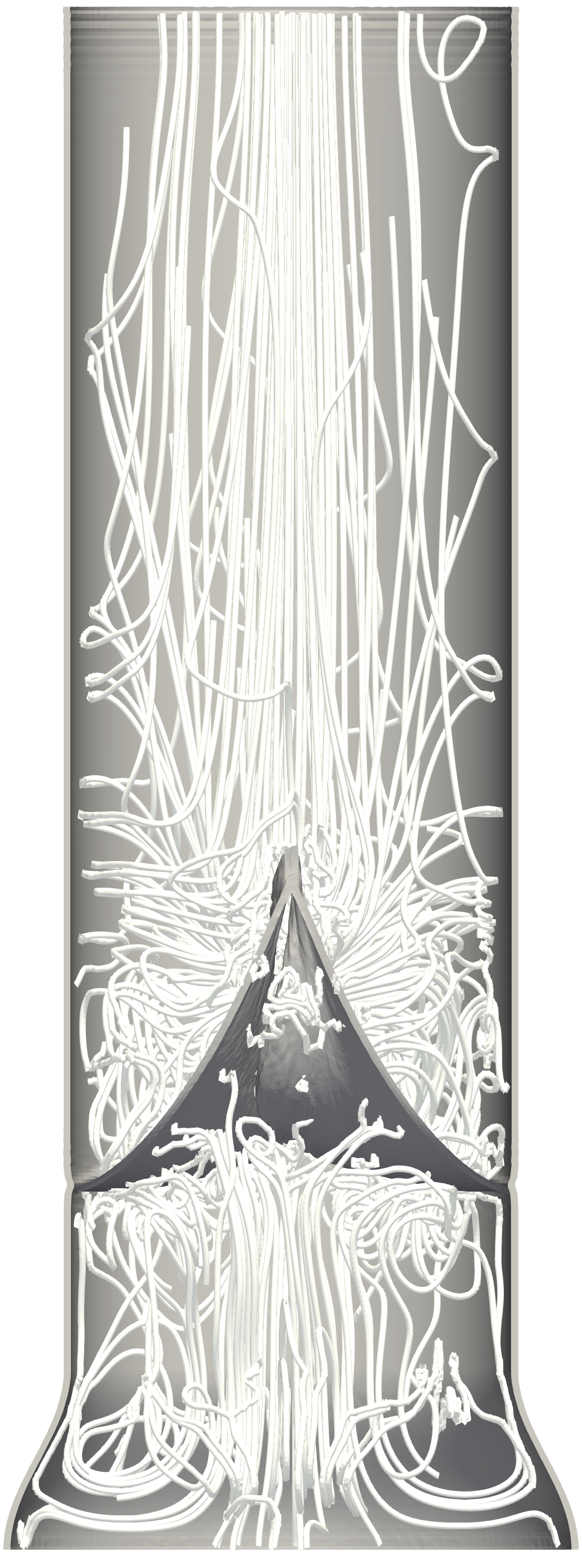} & 
\includegraphics[width=.10\textwidth]{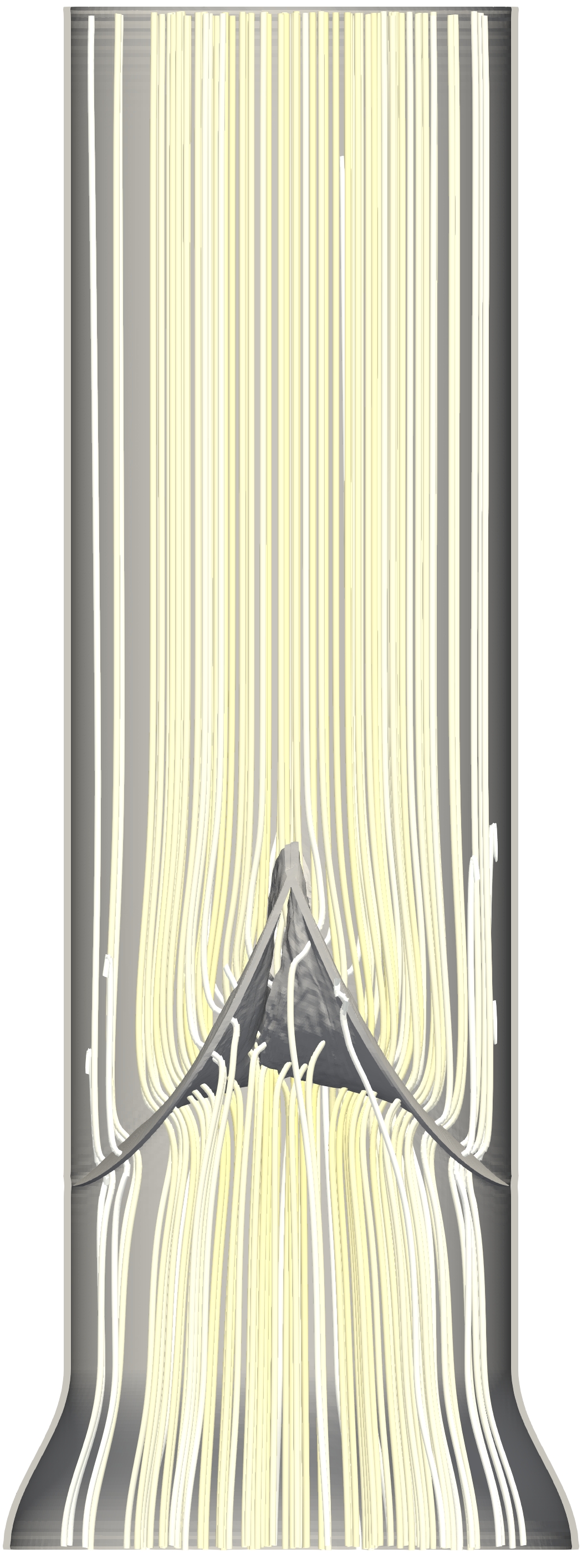} &
\includegraphics[width=.10\textwidth]{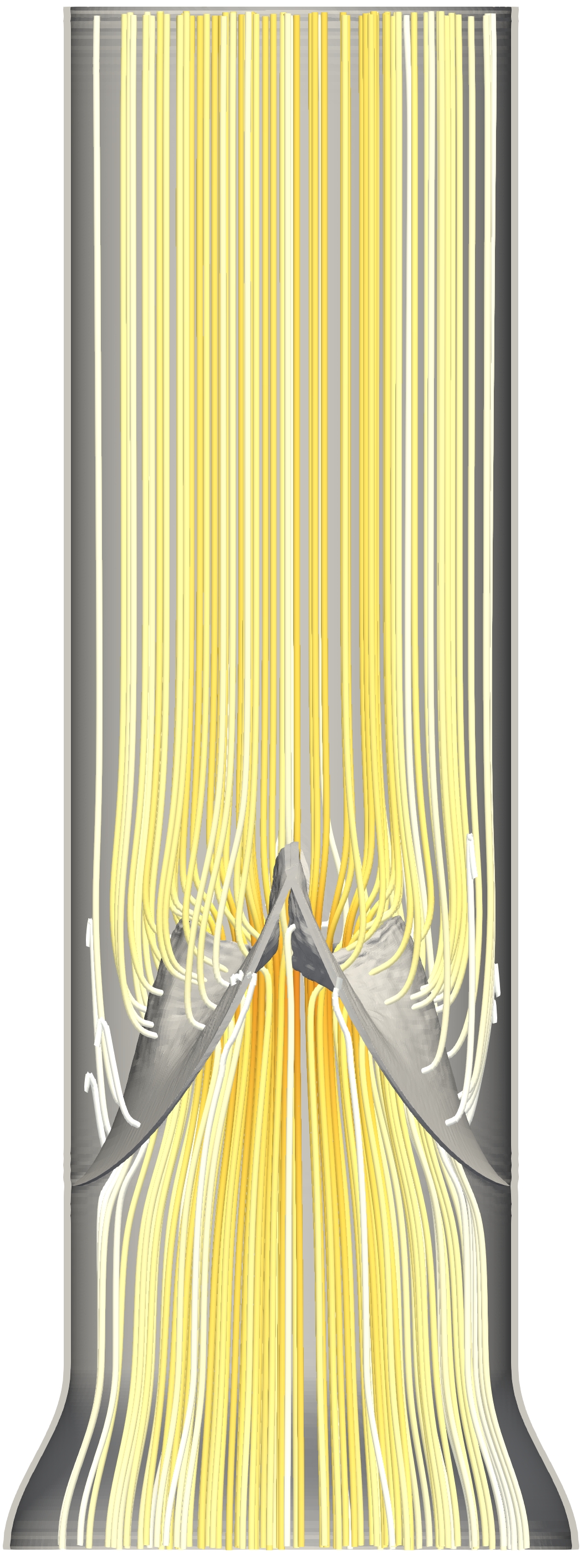} & 
\includegraphics[width=.10\textwidth]{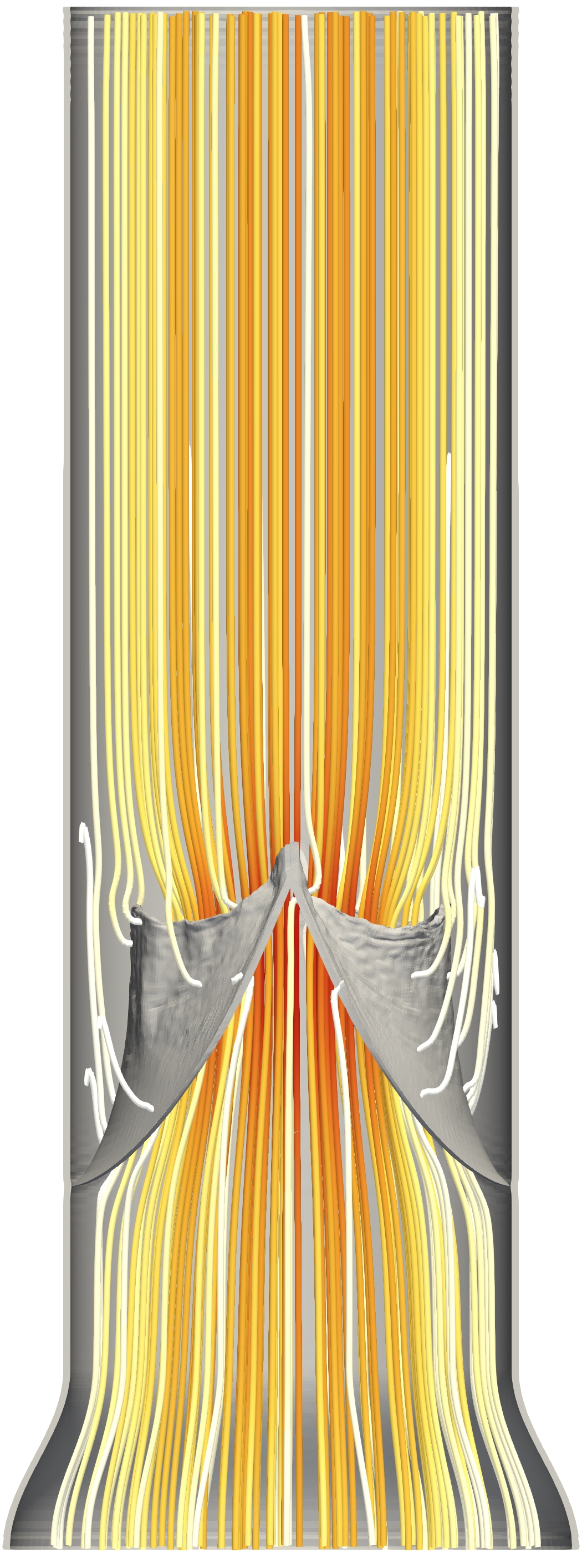} &
\includegraphics[width=.10\textwidth]{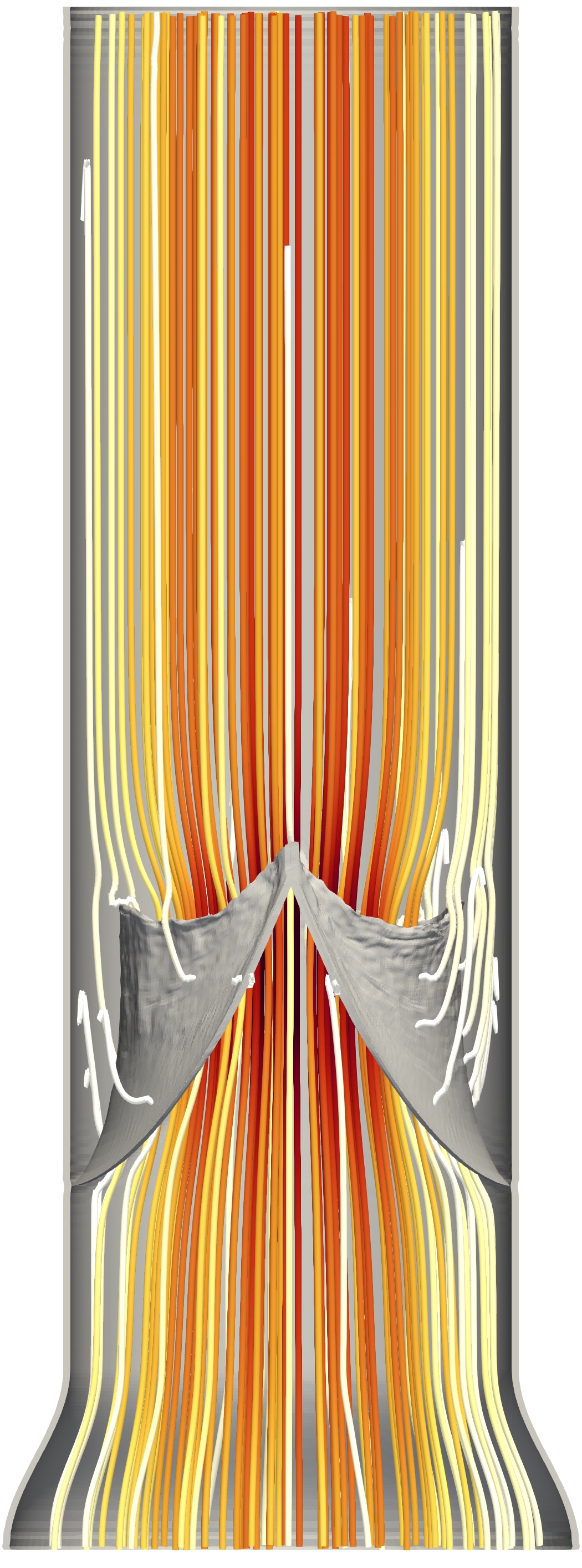} &
\includegraphics[width=.10\textwidth]{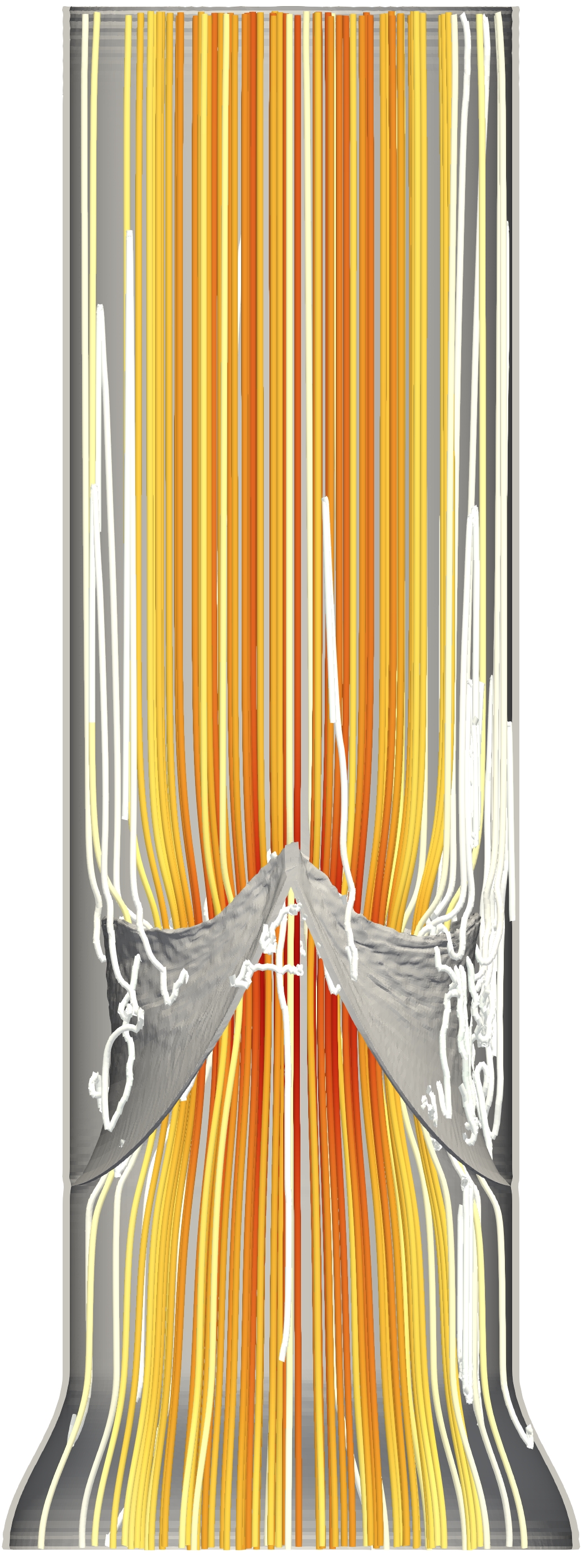} &
\includegraphics[width=.10\textwidth]{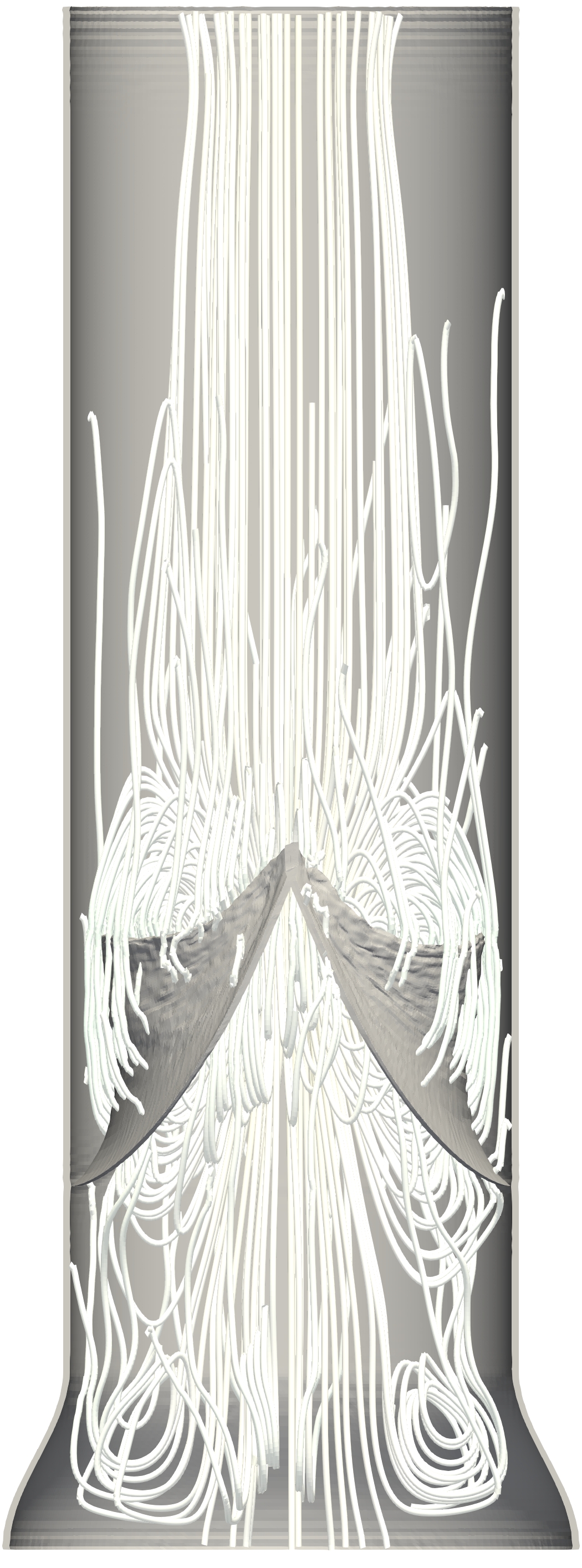} & 
\includegraphics[width=.10\textwidth]{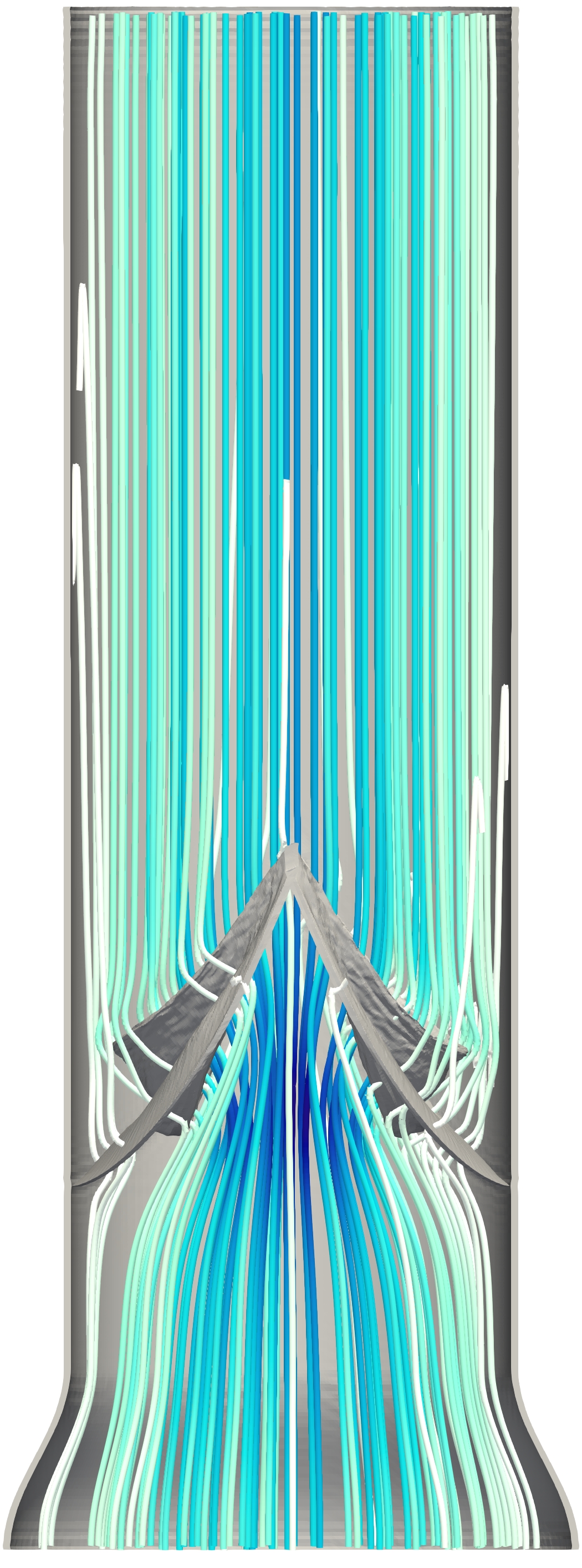} & 
\includegraphics[width=.10\textwidth]{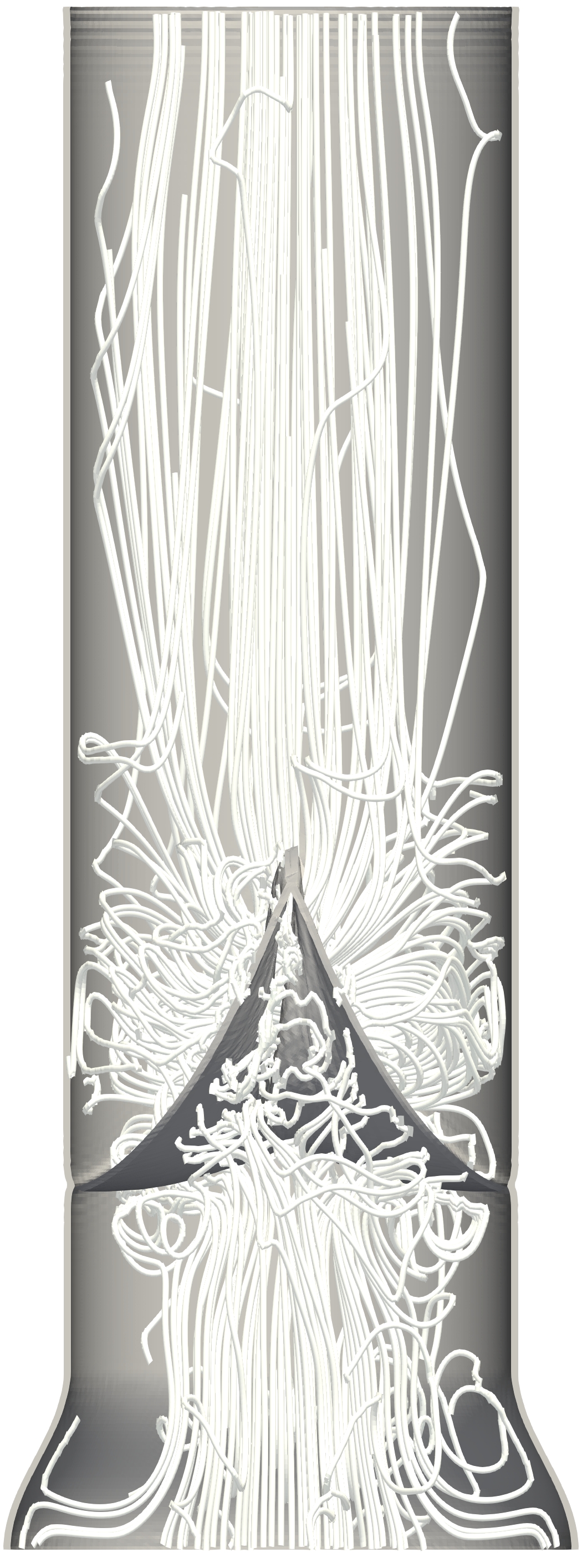} & 
\raisebox{13mm}{\includegraphics[width=.08\textwidth]{colorbar_zebrafish.jpeg}} \\ 
\\ 
\end{tabular}
\caption{Streamlines through the cardiac cycle.
The streamlines are colored by the vertical component of velocity to illuminate the flow direction.
}
\label{streamlines}
\end{figure}

Streamlines showed that the flow was laminar throughout the cardiac cycle (Fig. \ref{streamlines}). 
Streamlines during forward flow showed no evidence of vortex shedding from the leaflets with very little tangential motion. 
When the valve was opening or closing, the streamlines near the leaflet went through the leaflet, since the leaflet was moving with the local fluid velocity. 
In the neighborhood of the leaflets, the streamlines showed some tangential components with the opening and closing leaflet kinematics, but remained largely vertical away from the valve. 
While the valve was closed, the streamlines appeared less regular due to remaining flows of near-zero magnitude. 
Streamlines during the closing transient also remained laminar, and at no time was a major vortex visible immediately on the aortic side of the leaflets.

\begin{figure}[t]
\includegraphics[width=.49\textwidth]{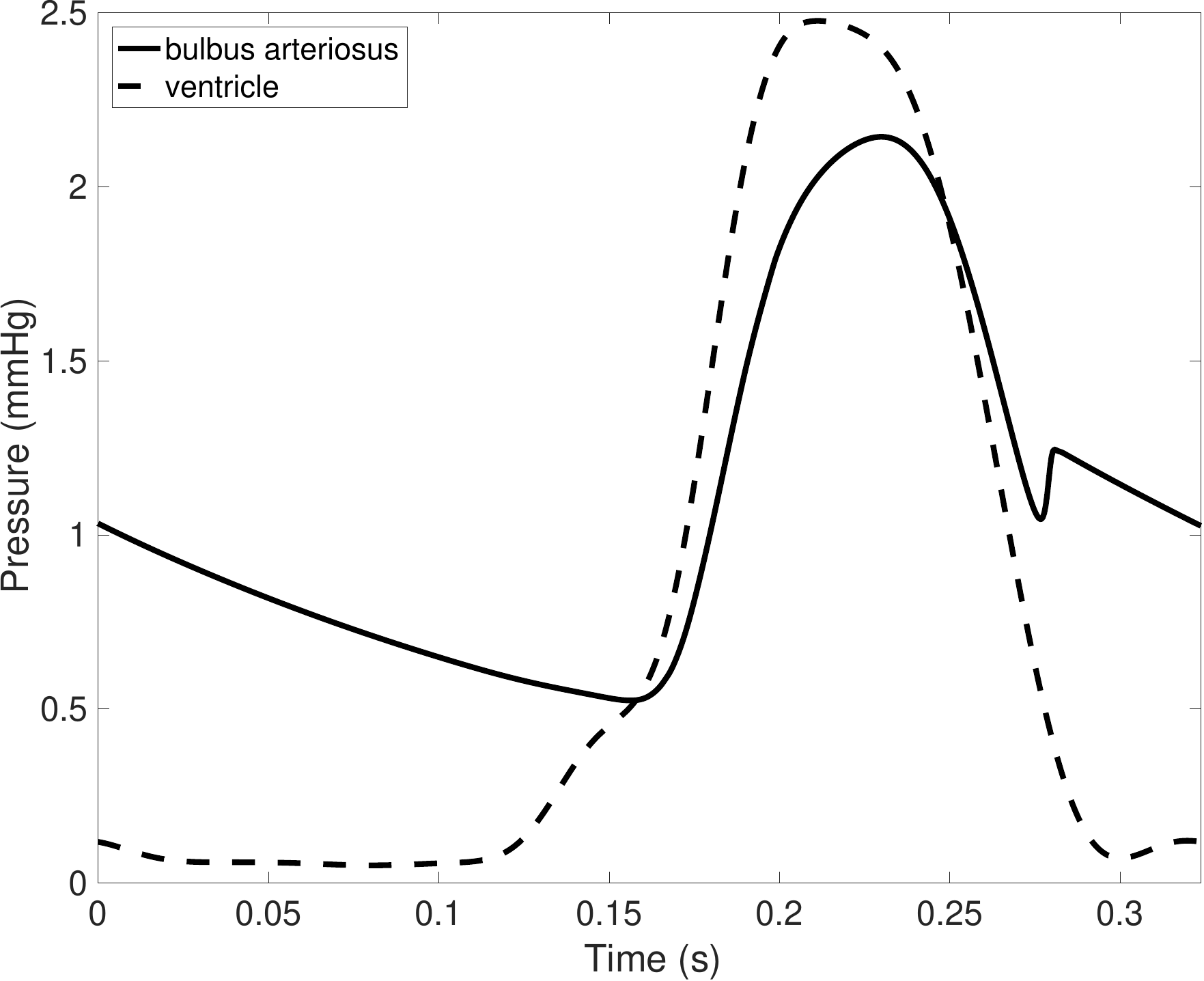}
\includegraphics[width=.49\textwidth]{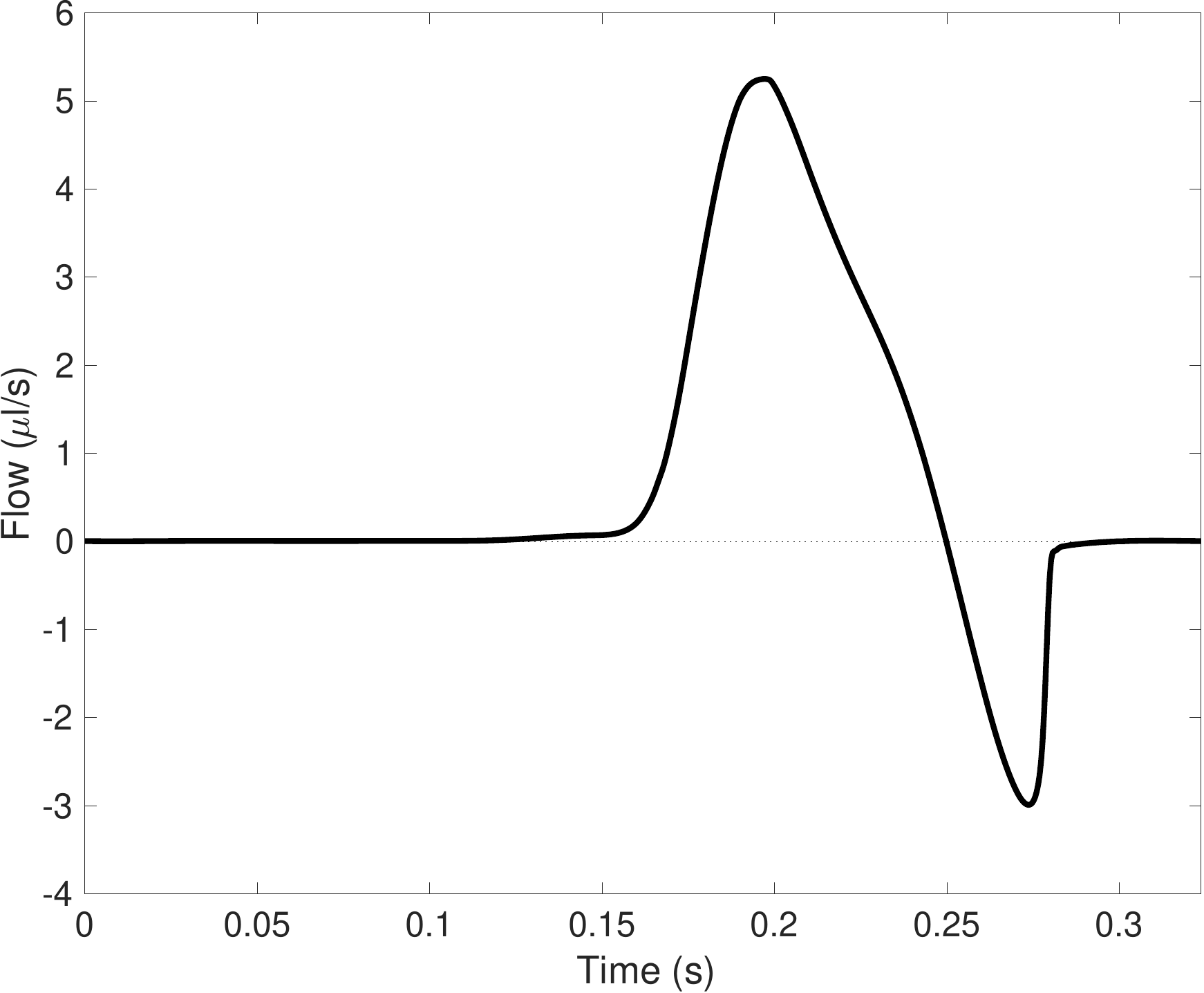}
\caption{Pressure and flow waveforms. }
\label{pressure_flow}
\end{figure}

Pressure and flow rate waveforms showed approximately physiological values \citep{hu2001cardiac,van2023fluid} through the cardiac cycle (Fig. \ref{pressure_flow}). 
The aortic pressure resulted from the coupled dynamics of the full FSI system and the RCR boundary conditions. 
The pressure lowered gradually through end diastole, rose with systole, showed a prominent dicrotic notch upon valve closure, then again began to decrease gradually during diastole. 
The aortic pressure took minimum, maximum and mean values of 0.52, 2.14 and 1.10 mmHg, respectively. 
Across the valve (and including the surrounding test chamber), the peak pressure gradient was 0.62 mmHg and mean gradient was 0.34 mmHg. 
The flow rate was near zero during diastole, then showed rapid increase with forward pressure, followed by sustained forward flow. 
After the pressure difference changed sign, the valve initiated closure. During this time, the flow rate was transiently negative as the blood that was between the closing leaflets moved toward the ventricle. However, this does not indicate regurgitation.
% the flow rate shows a transient of backflow.
%The majority of this was likely blood that has not fully passed the leaflet free edge. 
The valve returned to its closed state, and the flow rate settled around zero without regurgitation or leak.

The stroke volume was 266 nl and the maximum flow rate was $Q_{max} = $ 5.25 $\mu$l/s. 
The mean systolic flow was $Q_{mean}$ = 2.85 $\mu$l/s, which was computed between the time at which the pressure difference across the valve turned positive to the time at which the flow rate turned negative. 
The maximum Reynolds number was $Re_{max} = \rho (Q_{max}/A) r / \mu = 2.30$ and the mean Reynolds number was $Re_{mean} = \rho (Q_{mean}/A) r / \mu = 1.25$, where $A$ denotes the cross sectional area of the test chamber. 
The Reynolds number is thus order one and the flow lies in an intermediate regime, neither Stokes flow nor a fully inertial regime. 

\begin{figure}[t]
\setlength{\tabcolsep}{1.0pt}
\begin{tabular}{ | c | c | c | c | c | c |  }
% \begin{tabular}{ c  c  c  c  c  c  c  c  c}
% \hspace{-40pt}
\multicolumn{6}{c}{stretch} \\ 
\hline 
\multicolumn{2}{|c}{circumferential} & & \multicolumn{2}{c}{radial} & \\ 
\hline 
systole & diastole & & systole & diastole & \\ 
\hline 
\includegraphics[width=.22\textwidth]{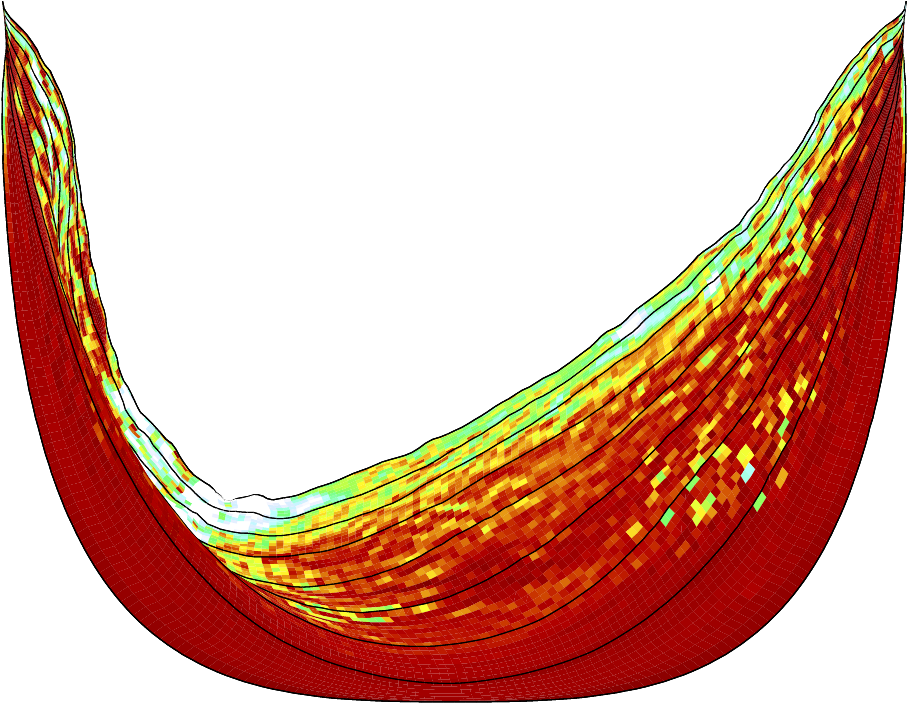} & 
\includegraphics[width=.22\textwidth]{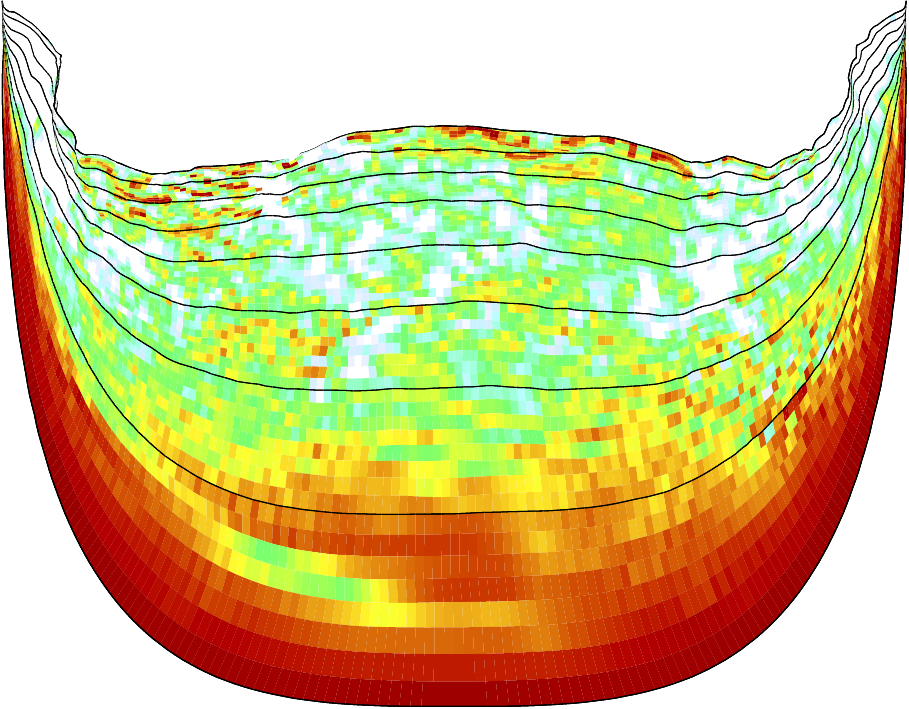} & 
\includegraphics[width=.038\textwidth]{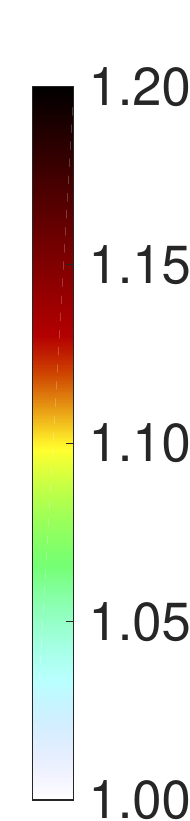} &
\includegraphics[width=.22\textwidth]{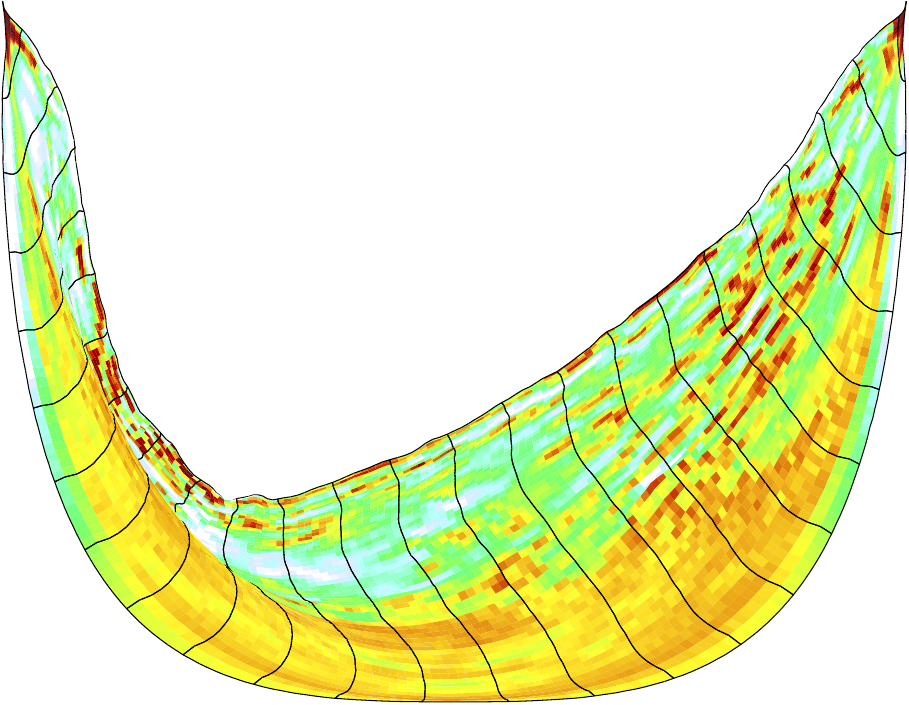} & 
\includegraphics[width=.22\textwidth]{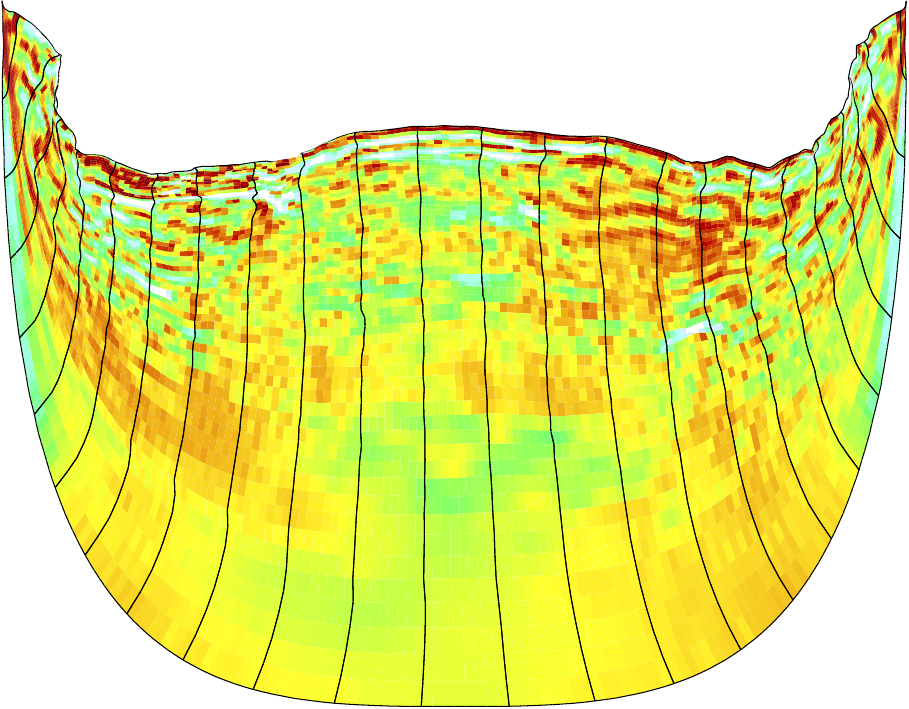} & 
\includegraphics[width=.038\textwidth]{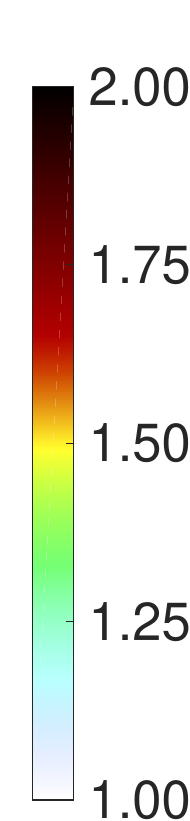} \\ 
\hline 
\multicolumn{6}{c}{} \\ 
\multicolumn{6}{c}{stress} \\ 
\hline 
\multicolumn{2}{|c}{circumferential} & & \multicolumn{2}{c}{radial} & \\ 
\hline 
systole & diastole & & systole & diastole & \\ 
\hline 
\includegraphics[width=.22\textwidth]{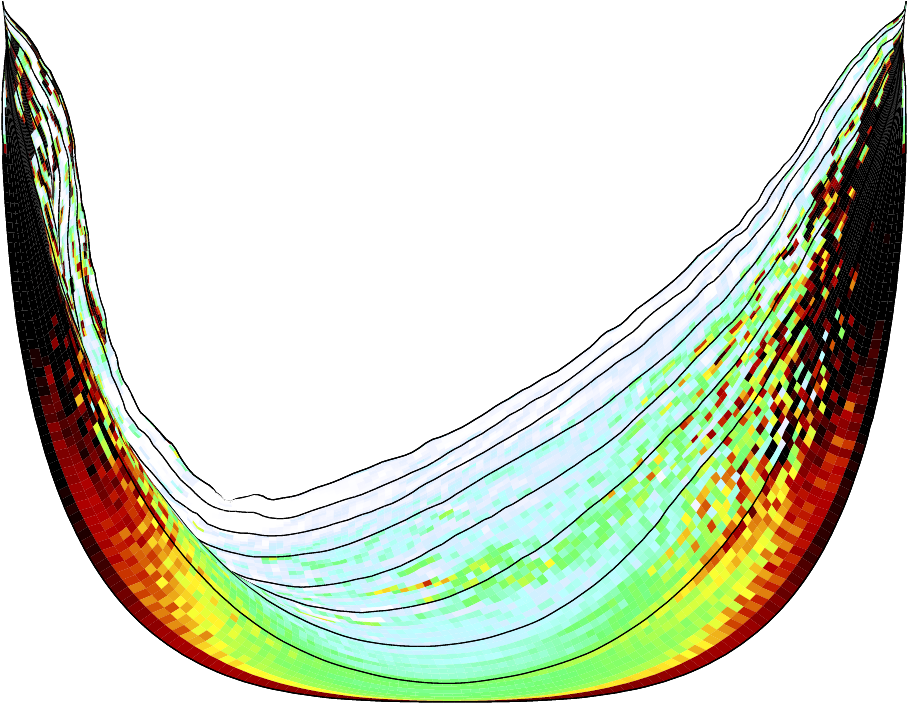} & 
\includegraphics[width=.22\textwidth]{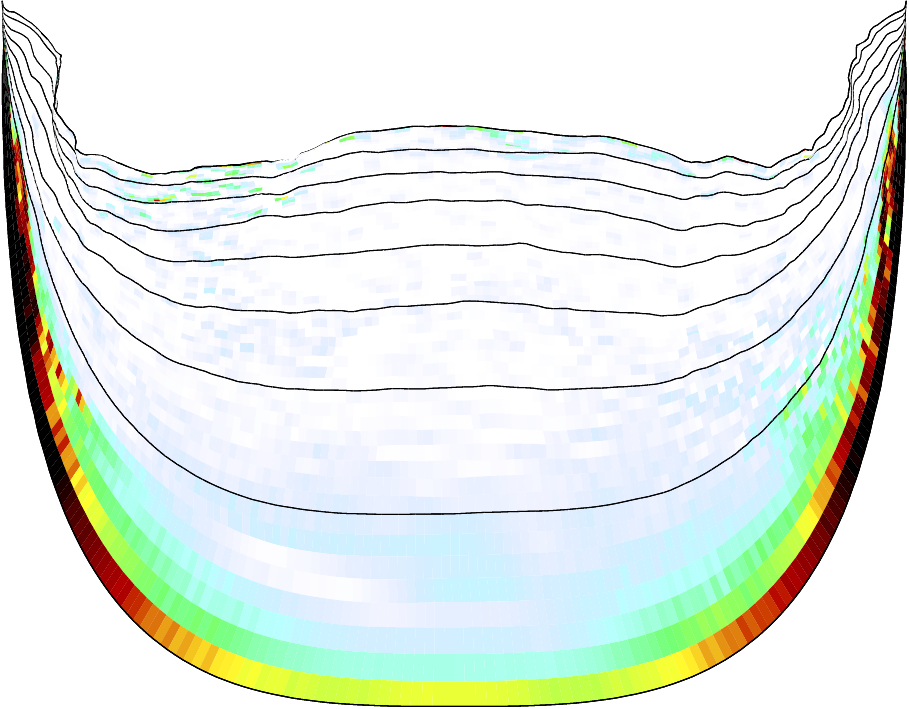} & 
\includegraphics[width=.038\textwidth]{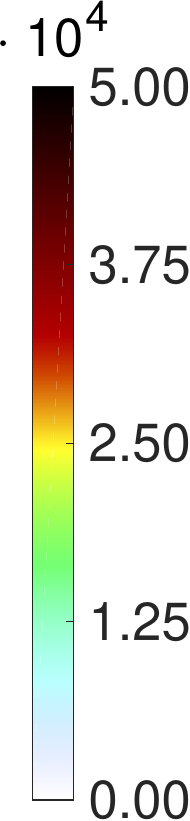} &
\includegraphics[width=.22\textwidth]{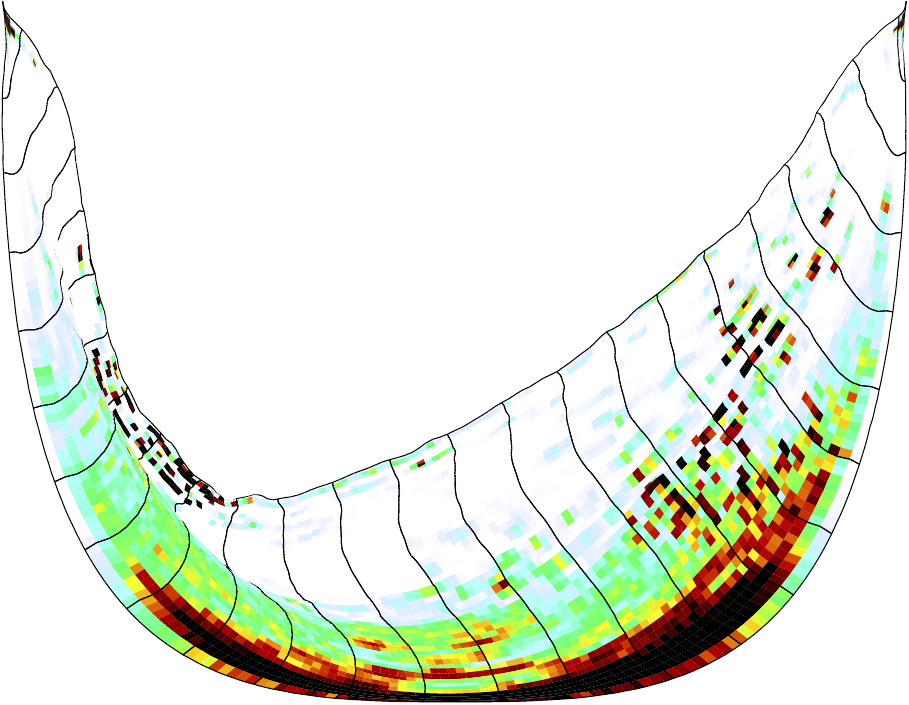} & 
\includegraphics[width=.22\textwidth]{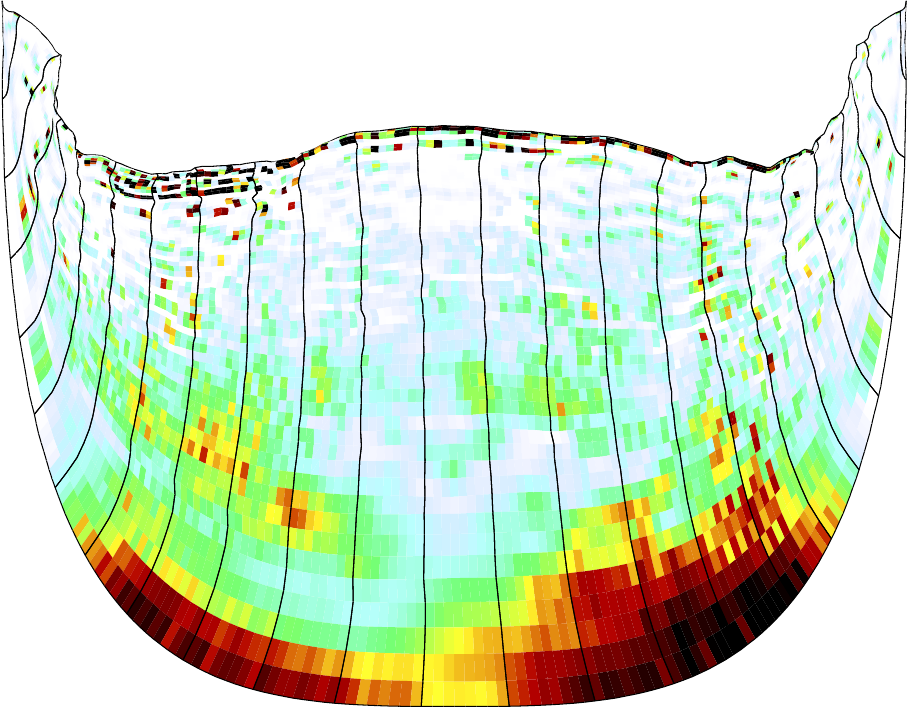} & 
\includegraphics[width=.038\textwidth]{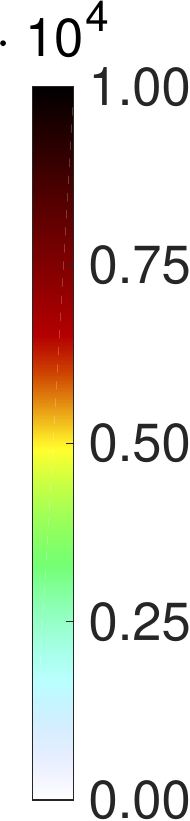} \\  
\hline 
\multicolumn{6}{c}{} \\ 
\multicolumn{6}{c}{tangent modulus} \\ 
\hline 
\multicolumn{2}{|c}{circumferential} & & \multicolumn{2}{c}{radial} & \\ 
\hline 
systole & diastole & & systole & diastole & \\ 
\hline 
\includegraphics[width=.22\textwidth]{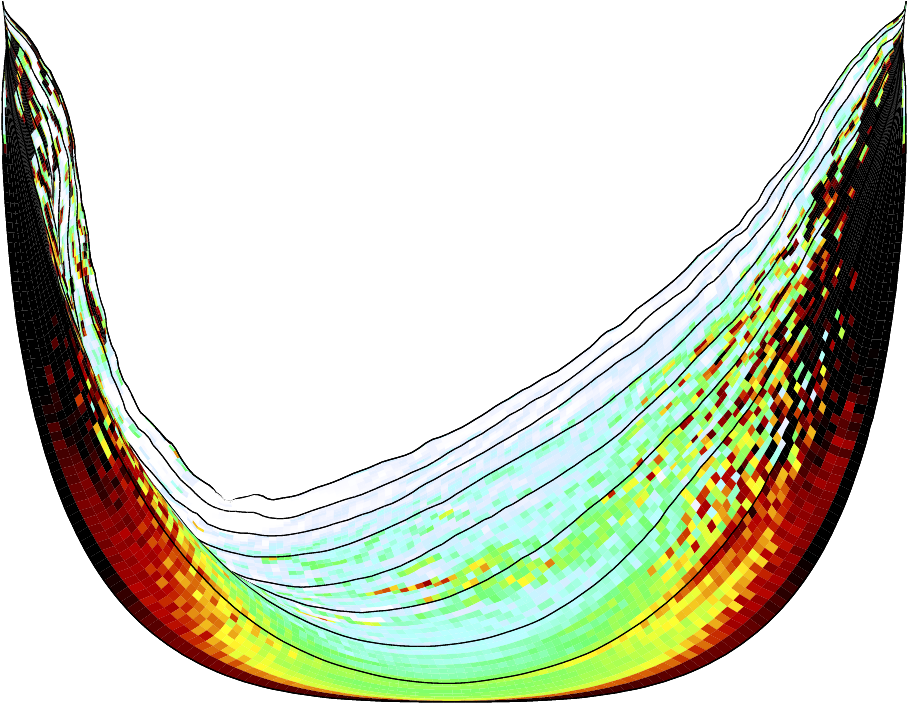} & 
\includegraphics[width=.22\textwidth]{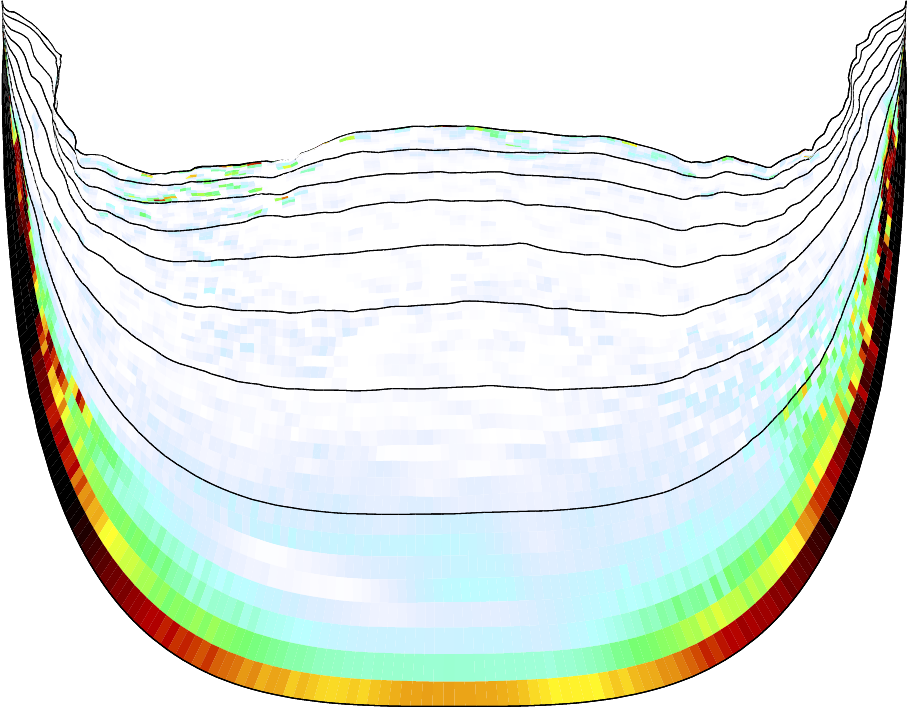} & 
\includegraphics[width=.038\textwidth]{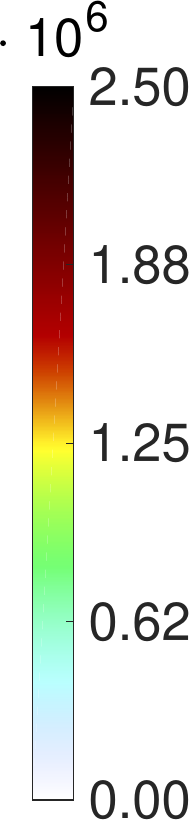} &
\includegraphics[width=.22\textwidth]{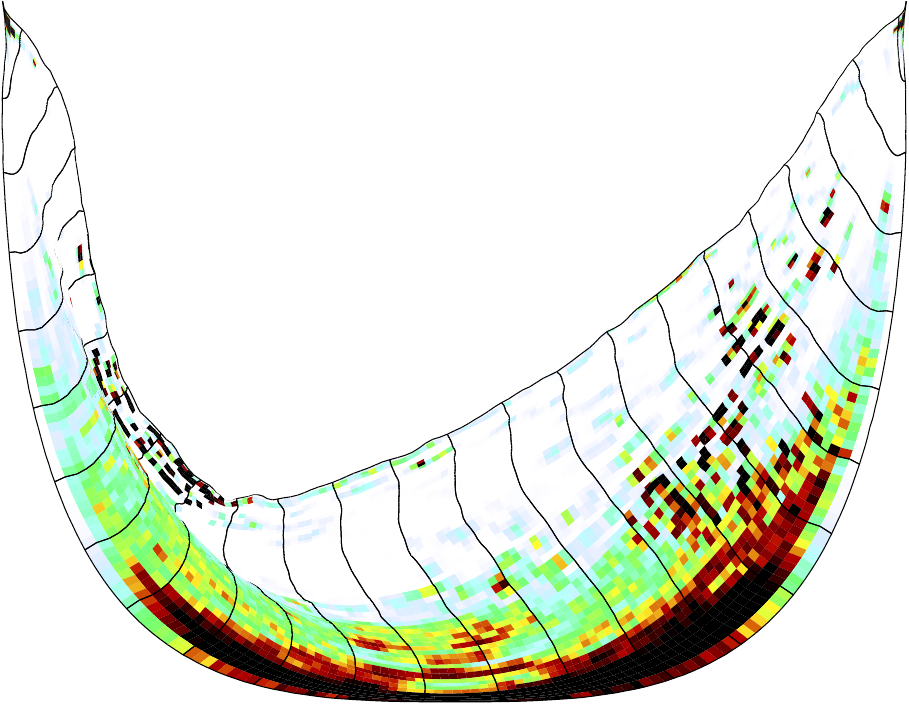} & 
\includegraphics[width=.22\textwidth]{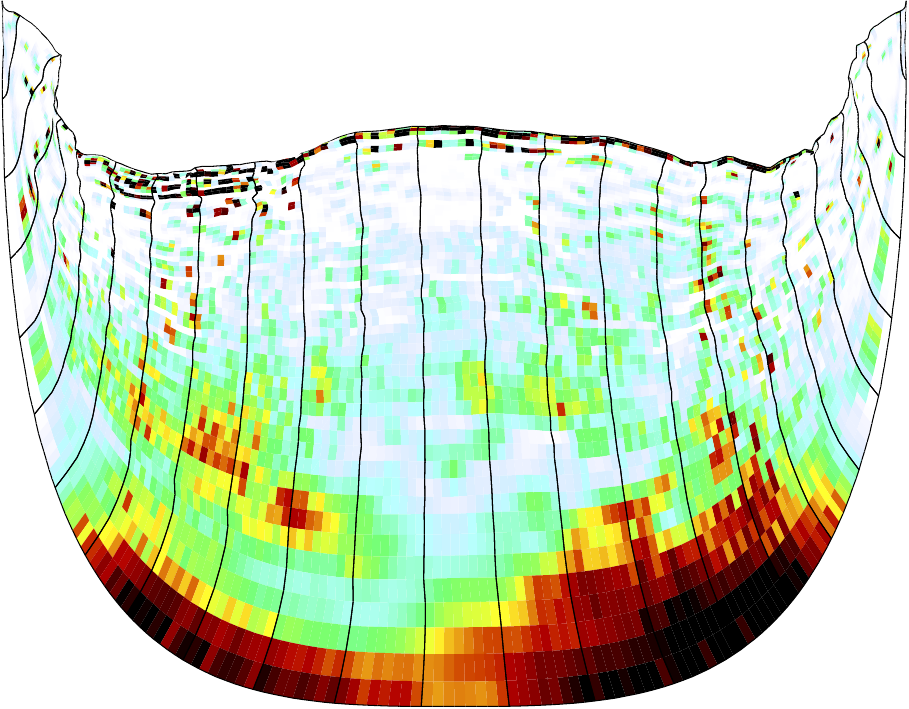} & 
\includegraphics[width=.038\textwidth]{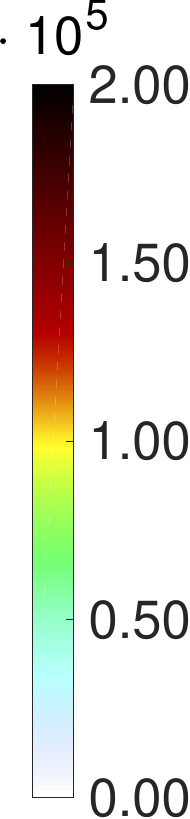} \\ 
\hline 
\end{tabular}
\caption{Stretch, stress (dynes/cm$^{2}$) and tangent modulus (dynes/cm$^{2}$) circumferentially and radially in systole and diastole.
Note differing scales throughout the figure.}
\label{stretch_plots}
\end{figure}

The stretch, stress and tangent modulus were anisotropic and generally higher in diastole when the valve is closed (Fig. \ref{stretch_plots}).
In diastole, the circumferential stretch was near to the predicted loaded stretch, which is expected by construction, away from the free edge. 
The stretch was lower in the coapted region where the valve is presumably less loaded circumferentially.
The stress and tangent modulus showed a concentration of stress and tangent modulus near the commissure. 
Stresses were more modest in the belly of the leaflet, and lower still at the free edge. 
The radial stretch, stress and tangent modulus were higher near the annulus and lower towards the free edge. 
The mean values of circumferential stretch, stress and tangent modulus were 1.12, $2.36 \cdot 10^{4}$ dynes/cm$^{2}$ and $1.36 \cdot 10^{6}$ dynes/cm$^{2}$, respectively. 
The mean values of radial stretch, stress and tangent modulus were 1.44, $4.09 \cdot 10^{3}$ dynes/cm$^{2}$ and $9.16 \cdot 10^{4}$ dynes/cm$^{2}$, respectively.
The mean stretch in both directions was lower than the predicted stretch at which the models were designed, likely because the coaptation region contacts the opposing leaflet and is less loaded in this region. 
Accordingly, the mean tangent moduli were also modestly lower. 

In systole, the circumferential stretch, stress and tangent modulus are much lower than in diastole, as expected with much less pressure loading on the valve. 
The stretch was higher near the annulus and decreased towards the free edge, and stress and tangent modulus were correspondingly lower.  
Radially, more stretch remained during forward flow, likely due to axially aligned shear on the valve.  
Elevated radial stress and tangent modulus were observed near the annulus, with more modest stress and tangent modulus through the leaflet belly.

\section{Discussion and Conclusions}

In this work, we have constructed a model of the zebrafish aortic valve using a first principles approach. 
These methods allowed for estimates of material properties of zebrafish valvular tissue that are largely unknown. 
In FSI simulations under physiological pressures, the model showed realistic kinematics and flow rates. 
These appear to be the first three-dimensional FSI simulations of the zebrafish aortic valve.

This work demonstrates the effectiveness and validity of the design-based elasticity concept. 
In the zebrafish, velocities, pressure and length scales are very different from those used in previous humans studies. 
Despite these differences, the valve's physiological function suggests that the estimated stresses and tangent moduli are realistic. 
That design-based elasticity could be applied to the zebrafish, in the absence of experimental data that would typically be required to conduct mechanics simulations, demonstrates that it is a powerful tool for modeling in biomechanics. 
These methods are especially useful in the presence of limited data and potentially applicable to a broad set of problems.

This work shows interesting scalings and similarities compared to human values. 
The cardiac cycle duration is less than one order of magnitude shorter than that of humans. 
Approximately, 
valve diameter is lower by two orders of magnitude, 
pressure is lower by two orders of magnitude, 
velocity is lower by one order of magnitude,
stroke volume and flow rate are lower by five orders of magnitude, 
and Reynolds number is lower by three orders of magnitude, compared to corresponding values in humans. 
This lower Reynolds number resulted in highly laminar flow patterns with no vortices visible in the velocity field, in start contrast to human flows that show inertial effects including unsteady flow, vortices, and small regions of recirculation (even in the trileaflet case)  \citep{kaiser2022controlled}.

The leaflet kinematics also showed similarities, with some notable contrasts, to human studies.  
The qualitative kinematics were strikingly similar to those we observed in previous simulations, with similar appearance of the open configuration and orifice, as well as the closed configuration including appearance of slight excess free edge tissue  \citep{kaiser2023simulation}.

One notable difference was the closing transient, which showed a single negative oscillation, rather than a sustained oscillation with multiple sign changes in flow rate, that appeared proportionally longer in duration than that we observed in previous human simulation studies. 
Doppler echocardiography has detected outflow tract valve regurgitation in some adult zebrafish, both in wild-type individuals and in alk5a-/- mutants \citep{bensimon2022integration}.
Qualitatively similar closing transients in flow rate without regurgitation to those observed in this work have been observed in studies on dogs \citep{laniado1976hemodynamic}.
It has been hypothesized and shown that vortices shed from the leaflet and in the aortic sinus assist in human aortic valve closure \citep{bellhouse1969fluid}.
Given the Reynolds number and lack of vortices or a sinus, this mechanism does not occur in our simulations of the zebrafish aortic valve, but the valve closes without leak nonetheless. 
This work thus suggests that while vortices may assist closure, this mechanism is not necessary for valve closure. 
The vortex, however, may shorten the closing transient or lower its magnitude.

Examining leaflet mechanics, experimental estimates on human valves predicted a tangent modulus of $9.9 \pm 1.8 \cdot 10^{7}$ dynes/cm$^{2}$ circumferentially and $2.3 \pm 0.4 \cdot 10^{7}$  dynes/cm$^{2}$ radially \citep{pham2017quantification}. 
Thus, the predicted tangent moduli in the zebrafish are two orders of magnitude lower that of humans. 
Other studies showed qualitatively similar trends to our results. 
A combined experimental/simulation study showed elevated stress at the commissures and somewhat elevated stress in the leaflet belly \citep{balguid2008stress}.
Another combined experimental/simulation study showed elevated radial stretch near the annulus, decreasing into the leaflet belly and lower still near the free edge, and elevated circumferential stretch and in the leaflet belly \citep{rego2022patient}.

The anisotropic stress field in mammalian aortic valve arises from its unique extracellular matrix (ECM) architecture that comprises of 3 layers: a fibrosa layer rich in circumferential collagen fibers facing the aortic side, a spongiosa middle layer rich in proteoglycans, and a ventricularis layer rich in radial elastin fibers facing the ventricular side \citep{billiar2000biaxial,gould2013hemodynamic,kodigepalli2020biology}. Moreover, the fibers near the commissure have the largest diameters and the lowest density, while the fibers near the fixed edge of mid-annulus are thin and high in density \citep{balguid2008stress}. Although a direct mechanical characterization of zebrafish aortic valves has not been possible, histological analysis indicates that they contain a similar elastin- and collagen-rich ventricularis-like layer and a proteoglycan-rich spongiosa-like layer \citep{schulz2019non}. However, the fibrosa layer does not seem to exist, and the leaflets are much more cellular compared to the mammalian leaflets. As the spatial organization of ECM componenets plays a critical role in valvular functions, our method provides a useful tool for researchers to assess the structural and mechanical remodeling of leaflets when studying zebrafish disease models \citep{chen2011cell}.

This study has limitations, in particular we lack experimental measurements of material properties and \emph{in vivo} valve geometry to validate our predictions. 
The test chamber is a cylindrical shape, rather than a beating heart or anatomical shape due to lack of \emph{in vivo} images at this developmental timepoint. 
We relied on an \emph{ex vivo} image of the zebrafish heart that was collapsed without pressure for estimating annular radius and height. 
In future work, \emph{in vivo} geometry of the valve and outflow tract could be incorporated as could heterogeneous or three-dimensional, non-membranous thickness of the valve leaflets. 
The application of similar methods to thick structures that cannot be viewed as membranous or structures with heterogeneous thickness is considerably more complex, due to equations \eqref{eq_eqns} becoming three dimensional and likely having a more complex pattern of strain.

To conclude, our modeling methods produce model geometry, fiber structure and material properties from nearly first principles that showed realistic behavior in  FSI simulations. 
These methods can thus be utilized for further studies of heart valves and their surrounding hemodynamics in the zebrafish. 
These findings serve as a basis for future studies on valve development and function in a commonly used model of cardiac development.

\section{Acknowledgements}

All authors were supported in part by NIH grants 5R01HL129727-08 and 5R01HL159970-04. 
EZ was supported in part by NIH grant T32HL144449.
Computing for this project was performed on the Stanford University's Sherlock cluster with assistance from the Stanford Research Computing Center. 
We would like to thank David Traver (UCSD) and Nathan Lawson (University of Massachusetts Medical School) for providing the fish line.

\bibliography{zebrafish_valve_refs}

\section{Appendix}
\label{appendix}

\subsection{Fluid-structure interaction details}

The FSI simulation setup is described here in detail. 
Fluid-structure interaction simulations were performed with the Immersed Boundary Method \citep{ib_acta_numerica} in IBAMR (Immersed Boundary Adaptive Mesh Refinement) \citep{griffith2010parallel}.
The fluid had a density of 1.0 g/ml and was modeled as Newtonian with viscosity 0.04 poise. 
The fluid mesh was a uniform staggered grid with 104 x 104 x 256 points and mesh width of 5 microns. This mesh size was selected following a mesh resolution study (Section \ref{resolution}).
The time step was set to $5 \cdot 10^{-7}$ s, which is highly restrictive and was required due to explicit coupling between fluid and structure, very small spatial resolution and only modestly lower material stiffness compared to human data. 
Simulations were run with 48 cores on Stanford University's Sherlock cluster and took approximately 2 days per cardiac cycle and 4 days total.

The valve was placed into a cylindrical tube of length 1.28 mm with diameter 363 microns with a flow extender at the inlet of length 150 microns and an additional radius 50 microns.  
The tube was meshed in cylindrical coordinates with edge length approximately 3 microns. 
The structure was held together by linear springs in the $r,\theta$ and $z$ directions, which exerted force as $f = k(L-R)/R$ with length $L$, rest length $R$ and relative spring constant $k = 2.56 \cdot 10^{-2}$ dynes. 
The tube included three layers each spaced half the fluid mesh width apart for a total thickness of 10 microns, serving to mitigate the ``grid-aligned artifact'' as in the leaflets. 
The tube was held in a fixed location with ``target points,'' or linear springs of zero rest length and stiffness 0.90 dynes/cm, which was selected to achieve minimal movement without affecting the time step restrictions. 
The additional radius of 50 microns served to stabilize inflow where the tube intersects the fluid domain boundary at the interface of the Neumann (pressure) boundary at the inlet and zero Dirichlet boundary conditions exterior to the vessel. 
It was found that with a straight inlet geometry, a non-physical, local increase in inflow can occur at this interface and grow during sustained inflow.
A smooth, constricting inflow geometry completely removed these spurious flows. 
A systematic study of this phenomenon will be left for future work.

The cardiac cycle duration was 0.32 s, or approximately 185 beats per minute \citep{hu2001cardiac} and simulations were run for two cycles. 
The first cycle was discarded due to initialization effects and results from the second cycle are shown. 
Times are reported from the start of the second cycle. 
The ventricular pressure was prescribed at the upstream inlet based on experimental measurements \citep{hu2001cardiac}, which were smoothed via convolution with a normalized cosine bump of radius .025 s and represented as a finite Fourier series with 600 frequencies. 

The aortic outlet pressure was governed via a RCR (resistor capacitor resistor) lumped parameter network that models the distal vasculature, following a brief initialization phase.  
The minimum, maximum and mean aortic pressure were estimated to be 0.89, 2.08 and 1.39 mmHg respectively \citep{hu2001cardiac}. 
The flow rate target was set to 266 nl \citep{van2023fluid}. 
The ratio of proximal to distal resistors was set to 0.065 via an experimental estimate in humans, as we lack a value of this ratio for the zebrafish \citep{laskey1990estimation}. 
The values were then estimated via the method previously described \citep{kaiser2020designbased}. 
This process resulted in values 
$R_{p} = 1.3 \cdot 10^{4}$ s dynes cm$^{-5}$, 
$R_{d} = 2.1\cdot 10^{6}$ s dynes cm$^{-5}$ 
and 
$C = 1.0 \cdot 10^{-7}$ cm$^{5}$  dynes$^{-1}$ 
for proximal resistance, distal resistance and capacitance, respectively. 

\subsection{Mesh resolution study}
\label{resolution}

To evaluate the effects of mesh resolution, a comparison with coarser resolution was performed. 
The coarse fluid mesh resolution was set to 10 microns with a time step of $2 \cdot 10^{-6}$ s. 
The coarse solid mesh was targeted to 5 microns in the leaflets and cylindrical tube.  

Precise convergence in the immersed boundary method is difficult to achieve, due to the finite with of fluid-structure coupling, which  leads to interactions up to 2.5 fluid points away from any structural point. 
All structures thus possessed a nonzero finite effective width, causing the effective numerical orifice area between any fixed structures to increase with increased resolution, limiting to the exact position of the structure, and increasing flow rates accordingly. 
Further, since flow rates are directly coupled with the lumped parameter boundary conditions, precise periodic convergence is challenging to achieve in a coupled 3D/0D system.

\begin{figure}[th]

(a)
\begin{center}
\setlength{\tabcolsep}{0.0pt}
\begin{tabular}{ | c | c | c | c | c }
\cline{1-4}
\multicolumn{2}{|c|}{diastole} &  \multicolumn{2}{c|}{systole} \\ 
\cline{1-4}
coarse & fine & coarse & fine \\ 
\cline{1-4}
\includegraphics[width=.15\textwidth]{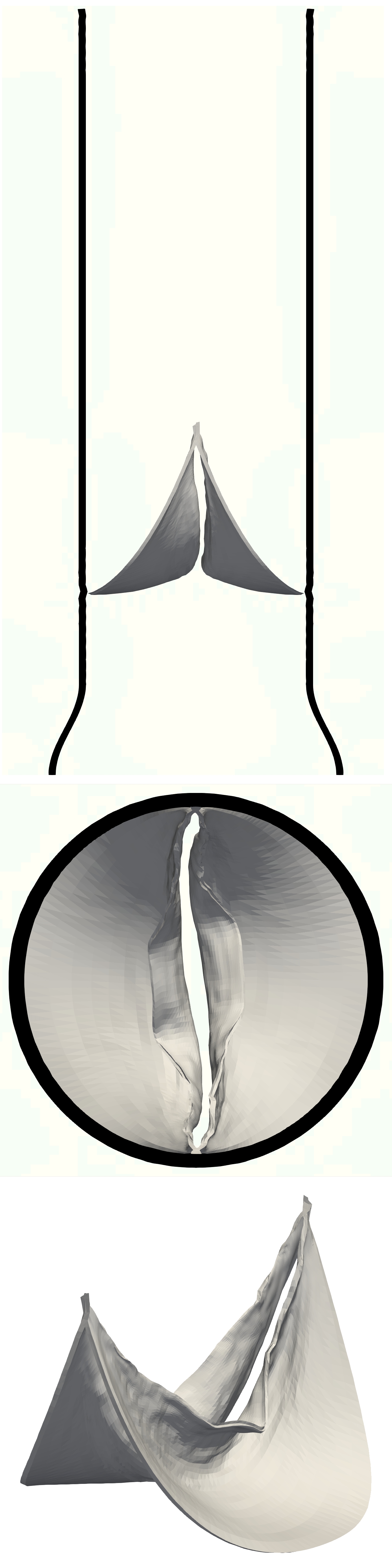} & 
\includegraphics[width=.15\textwidth]{aortic_50137444_384_0b73965_fish_tube_c_1pt57_r_pt017_init_interp_p1pt5_paraview_vertical0197.jpeg} & 
\includegraphics[width=.15\textwidth]{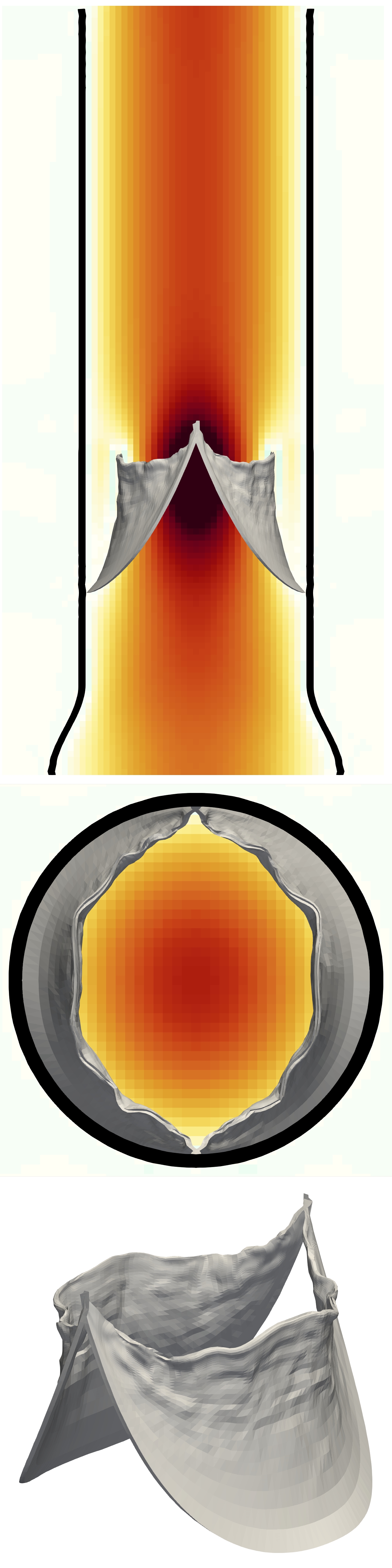} & 
\includegraphics[width=.15\textwidth]{aortic_50137444_384_0b73965_fish_tube_c_1pt57_r_pt017_init_interp_p1pt5_paraview_vertical0305.jpeg} & 
\raisebox{65mm}{\includegraphics[width=.08\textwidth]{colorbar_zebrafish.jpeg}}
\\ 
\cline{1-4}
\end{tabular}
\end{center}

(b)
\begin{center}
\setlength{\tabcolsep}{1.0pt}
\begin{tabular}{ c c }
\includegraphics[width=.49\textwidth]{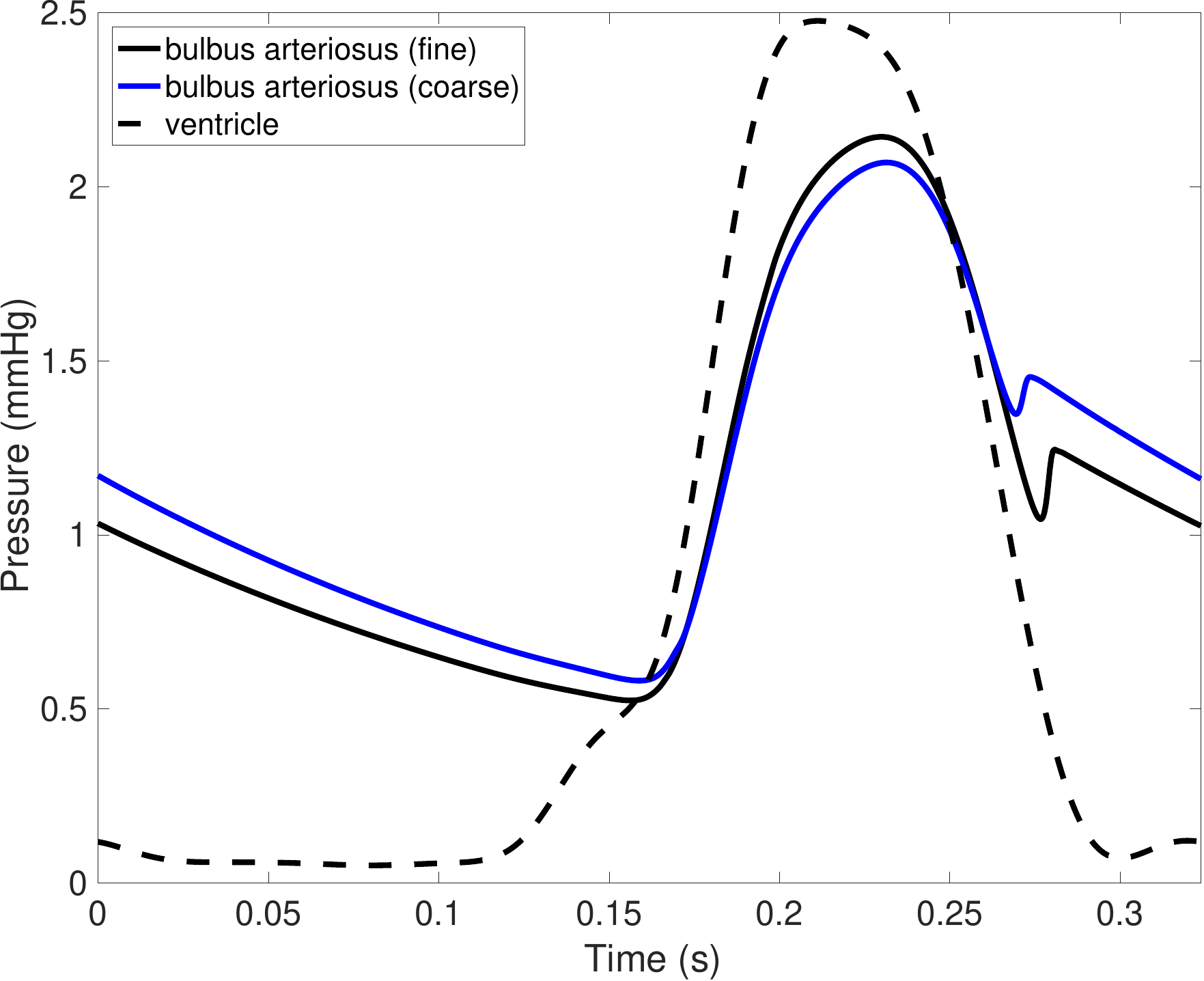} & 
\includegraphics[width=.49\textwidth]{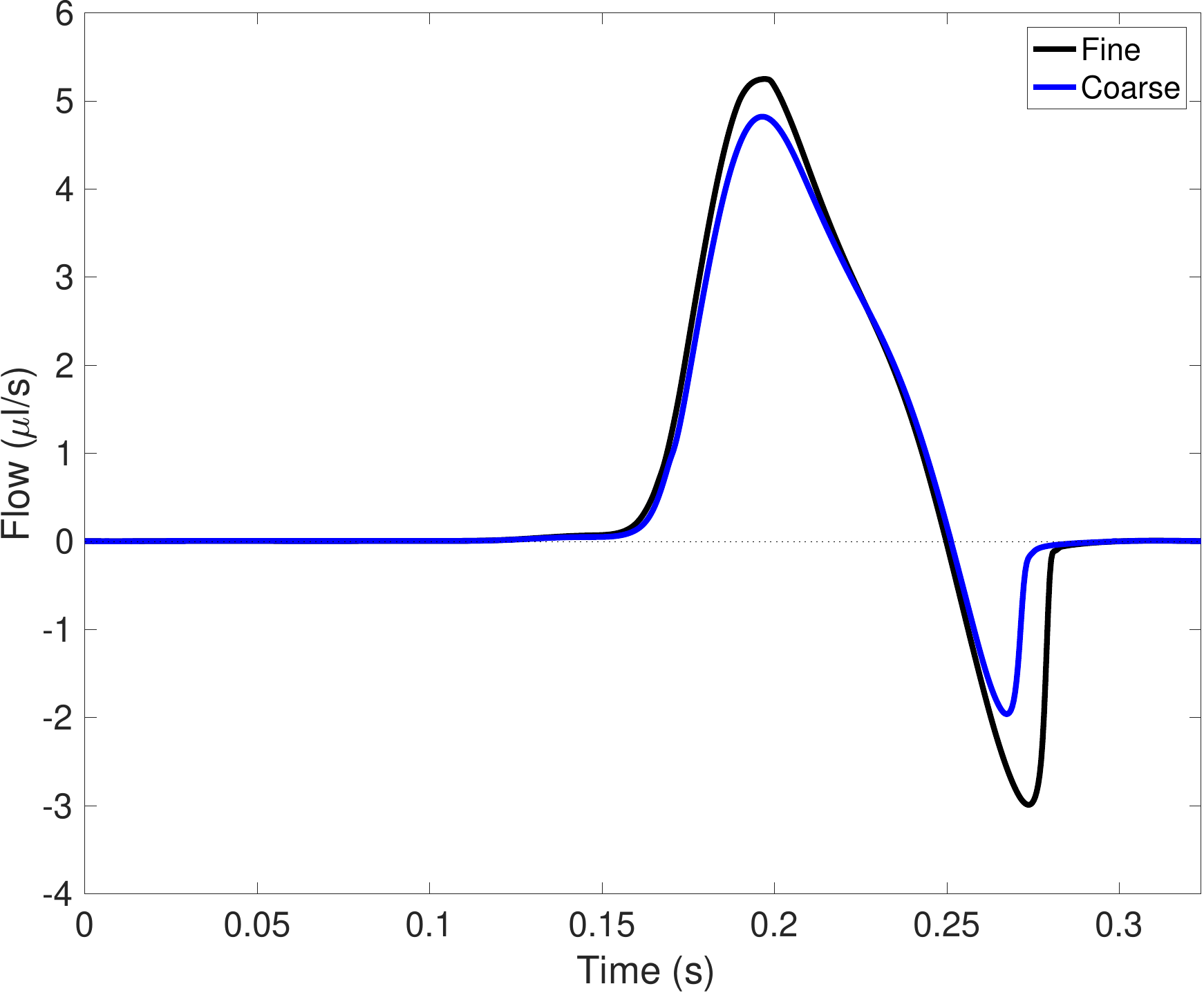}
\end{tabular}
\end{center}

\caption{(a) Comparison of coarse and fine resolution flow fields, which appear highly similar.
(b) Comparison of coarse and fine pressure and flow waveforms, which also are similar. 
Note that the ventricular pressure is prescribed and identical across both simulations. 
}
\label{appendix_flow}
\end{figure}

Despite these limitations, we observed qualitatively and quantitatively similar flow fields, flow rates and pressures between the coarse and fine simulations (Fig. \ref{appendix_flow}). 
The systolic flow field appeared similar in both cases, showing a smooth, laminar jet of similar magnitude without the appearance of vortices. 
The fine resolution shows slightly more leaflet excursion than that of the coarse resolution, caused by the increased implied width of the leaflets and wall in the coarse case. 
In closure, both flow fields are near zero and belly of the leaflets maintains a similar configuration. 
The precise position of the free edge showed differences between the two cases, likely to a physically unstable position of the lightly loaded free edge. 
The bulbus arteriosus pressures were similar in the two cases. 
The coarse flow rate was subtly lower than the fine flow rate, as expected because the increased with of the discrete delta function coupling created a smaller effective orifice area and thus lower flow rate. 

In conclusion, simulations were overall sufficiently similar between these two cases that coarsening resolution would not change any conclusions, nor would increasing resolution further be likely to alter conclusions. 
Doubling the spatial resolution to 2.5 microns would be extremely computationally expensive due to time step restrictions, which are expected to be 4 times lower than that of the current fine resolution or $1.25 \cdot 10^{-7}$ s or less. 
Due to similarities between 5 and 10 micron resolution, this would be unlikely to alter any conclusions. 
Results from fine resolution are presented throughout the main manuscript.

\end{document}